\documentclass[a4paper,11pt]{article}
\pdfoutput=1 

\usepackage{jheppub} 

\usepackage[T1]{fontenc} 
\usepackage{multirow}

\title{\boldmath Dark side of the Seesaw}

\author[a]{Subhaditya~Bhattacharya,}
\author[b,c]{Ivo~de~Medeiros~Varzielas,}
\author[a,d]{Biswajit~Karmakar,}
\author[c]{Stephen~F.~King,}
\author[a]{and Arunansu~Sil}

\affiliation[a]{Department of Physics, Indian Institute of Technology Guwahati,
Assam-781039, India}
\affiliation[b]{CFTP, Departamento de F\'{\i}sica, Instituto Superior 
T\'{e}cnico,Universidade de Lisboa, Avenida Rovisco Pais 1, 1049 Lisboa, 
Portugal}
\affiliation[c]{School of Physics and Astronomy, University of Southampton,
SO17 1BJ Southampton, United Kingdom}
\affiliation[d]{Theoretical Physics Division, Physical Research Laboratory,
 Ahmedabad-380009, India }

\emailAdd{subhab@iitg.ac.in}
\emailAdd{ivo.de@udo.edu}
\emailAdd{biswajit@prl.res.in}
\emailAdd{king@soton.ac.uk}
\emailAdd{asil@iitg.ac.in}

\abstract{In an attempt to unfold (if any) a possible connection between two apparently 
uncorrelated sectors, namely neutrino and dark matter, we consider the type-I 
seesaw and a fermion singlet dark matter to start with.  Our construction 
suggests that there exists a scalar field mediator between these two 
sectors whose vacuum expectation value not only generates the mass of the 
dark matter, but also takes part in the neutrino mass generation. While the 
choice of $Z_4$ symmetry allows us to establish the framework, the vacuum expectation value of the 
mediator field breaks $Z_4$ to a remnant $Z_2$, that is responsible to keep dark matter
stable. Therefore, the observed light neutrino masses and relic abundance 
constraint on the dark matter, allows us to predict the heavy seesaw scale as 
illustrated in this paper.The methodology to connect dark matter and neutrino 
sector, as introduced here, is a generic one and can be applied to other 
possible neutrino mass generation mechanism and different dark matter 
candidate(s).}

\begin{document} 
\maketitle
\flushbottom

\section{Introduction}

The hint of physics beyond the Standard Model (SM) has come from the measurement of
non zero neutrino masses and astrophysical observations supporting the existence of dark matter (DM). 
It is indeed intriguing to identify a common origin of both of these weakly coupled sectors.

In spite of earlier attempts to bring the dark and neutrino sector under one umbrella 
(see for example, \cite{nudmrefs}), one-to-one correspondence between the dark 
sector to a specific 
scenario beyond the SM responsible for seesaw mechanism to generate neutrino masses 
hasn't been firmly established. The main aim of this analysis is therefore to identify a simple common origin which initiates 
seesaw mechanism for neutrino mass generation and controls also the DM phenomenology. 
We point out that if the theory assumes the existence of an additional scalar singlet ($\phi$) which couples to both DM
and neutrino sector and assumes a non-zero vacuum expectation value (vev), it may generate 
light neutrino mass through seesaw of type-I~\cite{Minkowski:1977sc, 
Mohapatra:1979ia, GellMann:1980vs, Schechter:1980gr}, 
while also yield DM mass. Then, the observed light neutrino masses and the relic density constraint on DM 
(that crucially controls DM mass) can indicate a particular value of the seesaw scale (or a range of values), thus establishing 
a common origin of the neutrino and dark sectors. The challenge here is then to choose a symmetry that allows 
the interactions between dark and neutrino sector, and keep an unbroken symmetry intact to protect the DM from decaying 
into either neutrinos or to the SM particles after spontaneous symmetry breaking (SSB) through the vev of $\phi$. 
We demonstrate that the assumption of a $Z_4$ symmetry under which $\phi$ transforms as $2$ in additive notation (and 
suitable choices of the $Z_4$ charges of the other DM and SM particles) the connection can be securely established, while after 
SSB, the theory keeps a remnant $Z_2$ symmetry to stop the DM from decay. 
The dark sector phenomenology has been kept minimal; that is of a 
fermionic DM, in the form of a singlet Majorana fermion ($\chi$) coupled to the visible sector through the singlet scalar $\phi$.  
Thanks to the mixing of $\phi$ with SM Higgs due to SSB, the fermion DM can annihilate to SM particles and 
obtains a thermal freeze out. The seesaw mechanism is also chosen to be the simplest of its kind, type-I, 
assuming  light neutrino mass. 

The mechanism shown here is apparently the simplest of its kind, and a generic one; with many possible extensions either 
to relate other types of DM sector or to a different type of neutrino mass generation mechanism.  
The chosen set up allows a large region of parameter space where the constraints from dark sector 
and neutrino mass and oscillation data agrees together to indicate a limit on the seesaw scale. Correspondence between the
two sectors depend crucially how the DM mass is restricted 
from relic density and direct search constraints. Therefore, there is some model dependence in the
prediction of the Seesaw scale. The choice of the DM framework has partially been guided by the fact that 
the model is predictive and has a rather restrictive choices of DM mass possible from the relic and direct search data.
Collider search of this particular model is difficult due to absence of charged exotic final states. One has to depend instead 
on initial state radiation to recoil against the DM to yield the characteristic signature of jets plus missing energy. 
That however, goes well with non-observation of any excess in the missing energy channels studied at Large 
Hadron Collider (LHC) so far.    

The rest of the paper is organised as follows. We discuss the model 
and formalism in Section \ref{sec:model}. Scalar potential is discussed next 
with the interplay of Higgs mass and related observations in 
Section \ref{sec:scpot}. DM phenomenology comes next with relic density and 
direct search constraints in Section \ref{sec:dm}. In Section \ref{sec:conec}, 
we then discuss the allowed parameter space common to neutrino and dark matter 
sector to draw the connection. We finally summarise the outcome of the analysis 
in Section \ref{sec:conc}.
     
\section{The Model}\label{sec:model}

As stated previously, we choose the simplest seesaw extension of the SM 
with right handed neutrino ($N$), a minimal DM in the form of singlet Majorana fermion 
$\chi$ and a mediator singlet scalar field ($\phi$). We consider the existence 
of two sectors, 
visible and the hidden sectors. The visible sector has the usual SM field content. However, 
concerning our focus on the neutrino mass generation, the visible sector simplifies to the 
$SU(2)$ lepton doublets $L_i$ ($i$ being the generation index though we omit it 
in the rest of our discussion for simplicity) and SM Higgs doublet $H$. 
We also consider RH neutrino to be part of this sector. On the other hand, the mediator 
field $\phi$ and the DM field $\chi$ together forms the hidden sector. 
Note that all these additional fields (i.e. beyond SM fields) transforms 
non-trivially  under a $Z_4$ symmetry. The assignment of $Z_4$ charges to the fields is given in 
Table \ref{tab:1}. Note that while $L$ and $N$ carry Lepton numbers -1 and 1 respectively, 
the dark sector Majorana field $\chi$ does not have any Lepton number.  
\begin{figure}[h!]
$$
\includegraphics[height=2cm]{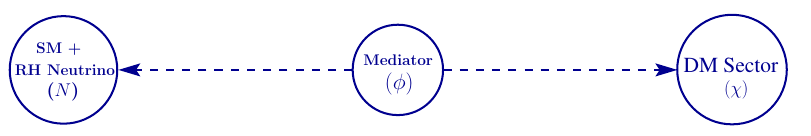}
$$
  \caption{Schematic representation for DM interaction with SM trough the 
scalar $\phi$ }\label{fig:schm}
\end{figure}
This minimal field content allows us to have a phenomenologically viable DM 
sector and type-I seesaw mechanism for neutrino mass generation (a toy model), 
which are connected by the vev of the mediator field. A schematic diagram of the framework is 
depicted in Fig. \ref{fig:schm}  for illustration purpose. Although, SM singlet 
scalars charged under additional symmetries acting as portals between the dark sector and SM 
have been explored in some earlier attempts \cite{Calibbi:2015sfa, Varzielas:2015sno, Bhattacharya:2016lts, 
Bhattacharya:2016rqj}, this has never been connected to the Yukawa neutrino sector to the 
best of our knowledge. 

\begin{table}[h]
\centering
\resizebox{5cm}{!}{%
\begin{tabular}{|c|ccccc|}
\hline
 Field &   $L$ & $H$ & $N$ & $\phi$ & $\chi$ \\
\hline
$Z_{4}$ & 0 & 0 & 2 &  2 & $1$\\
\hline
\end{tabular}
}\
\caption{\label{tab:1} {\small Transformation properties
fields involved under $Z_4$ in additive notation where charge $q$ means: the field transforms like $e^{i 2\pi q/4}$.}}
\end{table}

The Lagrangian of the framework can then be written as:
\begin{equation}\label{lag}
\mathcal{L}=\mathcal{L}_{SM}+ \frac{1}{\Lambda}\bar{L} \tilde{H} \phi N+
M_N \bar{N^c} N + y \phi \bar{\chi^c} \chi + V(H,\phi), 
\end{equation}
where $V(H,\phi)$ is the scalar potential involving the SM Higgs doublet $H$ and $\Lambda$ 
is the cut-off scale of the theory. 
We will discuss the details of the scalar potential and its implications below and in the next section.
Due to the $Z_4$ charges assigned, the only portal connecting visible and hidden
sectors is the non-renormalisable $\frac{1}{\Lambda}\bar{L} \tilde{H} \phi N$ term
as well as the Higgs portal couplings in the potential as discussed in the following section.
As $H$ (and $L$) are neutral, the charged lepton, up and down quark sectors are compatible 
with the neutrino sector, with the respective charges being neutral, their Yukawa couplings 
arise without coupling to $\phi$. Note that the model exhibits an 
effective theory approach to describe the neutrino sector. Although an underlying UV framework can 
be provided, the present set-up serves as an economic one in terms of keeping the fields 
content and symmetry minimal.

It is important to note that terms like $\bar{L} \tilde{H} N$, $\bar{N^c} N \phi$ are disallowed due to the $Z_4$ symmetry 
imposed on the model. The Majorana mass term for the DM fermion is 
disallowed by the $Z_4$ charges and will only be generated by the singlet scalar vev, through the allowed term $y \phi \bar{\chi^c} \chi$ involving the
fermion DM and the singlet scalar. Thus, dark matter 
mass $M_\chi \propto \langle \phi \rangle$. Additionally, according to the symmetries mentioned in 
Table \ref{tab:1}, a dim-5 term $\phi^2\bar{N^c} N/\Lambda$ is also allowed. However,  contribution 
of such a term in the effective light neutrino mass is small compared to the 
original contribution (with $M_N \gg \langle \phi \rangle^2/\Lambda$) and hence can be neglected. The vev of $\phi$
will also generate light neutrino masses{\footnote{A spontaneous breaking of a discrete symmetry 
may induce cosmological domain wall problem~\cite{Preskill:1991kd,Dvali:1994wv}. 
However this can be controlled provided a higher order discrete symmetry 
breaking term is present which does not affect our analysis.}} after the 
seesaw mechanism, suppressed by a factor of $\frac{\langle \phi \rangle^2}{\Lambda^2 }$,
\begin{equation}\label{mnu}
m_\nu \propto \frac{\langle \phi \rangle^2}{\Lambda^2 }
\frac{\langle H \rangle ^2}{ M_N},  
\end{equation}
yielding a correspondence between the DM mass ($M_{\chi} 
= y \langle \phi \rangle$) and the neutrino mass $m_{\nu}$ which depends also on the seesaw scale, $M_N$ and on the cut-off 
scale $\Lambda$.  The additional 
constraints of obtaining the correct relic abundance and direct search 
constraints for the DM, will control DM mass $M_\chi$ to a significant extent 
and therefore we can estimate a limit on the heavy seesaw scale, which is a main 
result of our analysis.

The salient features of the model are as follows: 
\begin{enumerate}
\item DM mass is generated through the vev of the mediator field, $\phi$.

\item Neutrino Yukawa interaction is allowed only with a dimension-5 operator involving 
the same mediator field $\phi$. 

\item The above two features of the model allow us to probe the seesaw scale, $M_N$, once 
the constraints from neutrino physics, DM relic density and direct detection results are incorporated. 
This however crucially depends on the choice of the cut-off scale $\Lambda$ of the theory. 
As we do not address a UV complete theory, $\Lambda$ is unknown. For the sake of simplicity 
and economy of parameters, we choose $\Lambda = M_N$ for illustration.
\end{enumerate}
\noindent A few comments before we analyse the model under consideration. Firstly, the correspondence 
between the dark sector and neutrino sector depends on how much one can restrict DM mass from relic density and direct search observation. The more 
the DM mass is relaxed, the less deterministic the heavy neutrino mass will be. The choice of the DM 
model has been motivated from above justification which we will elaborate shortly. 
Secondly, the $y$ coupling is restricted from perturbative limit to be $y \le \sqrt{4\pi}$. 
Although, in the subsequent analysis, $y$ has been replaced by  the ratio of the dark matter 
mass to the mediator vev ($\langle \phi\rangle=u$), $i.e.$ $y=\frac{M_\chi}{u}$, 
the limit turns out to be important in restricting the allowed parameter space of the model further, particularly 
that of the heavy dark matter mass regions. For example, choosing a cut-off scale of the theory as 
$\Lambda\sim 10^6$ GeV and demanding $y (\Lambda) \le \sqrt{4\pi}$, 
the limit on the coupling $y$ at the Electroweak scale turns out to be $y \le 1.16$. See 
Appendix A for details. 

\section{Scalar potential and heavy or light Higgs}\label{sec:scpot}

The complete scalar potential, involving $SU(2)_L$ doublet $H$ and singlet 
$\phi$ with the $Z_4$ charges as in Table 1, can be written
as~\cite{Robens:2015gla, Robens:2016xkb} 
\begin{equation}\label{eq:spot}
V(H,\phi)=-\mu_1^2H^{\dagger}H+\lambda_1(H^{\dagger}H)^2-\frac{1}{2}
\mu_2^2\phi^2+ 
\frac{1}{4}\lambda_2\phi^4+\frac{1}{2}\lambda_{12}\phi^2H^{\dagger}H. 
\end{equation}
We note here, that the coefficient of $\phi^2$ is deliberately chosen negative ($\mu_2^2>0$) so that 
it acquires a non-zero vev. The term  involving $\phi^2H^{\dagger}H$ yields mixing
between the scalars. In the unitary gauge we can write, $H=\frac{1}{\sqrt{2}}\left(
                                   \begin{array}{c}
                                     0\\
                                     \tilde{h}+v
                                    \end{array}
                                   \right)$,  and take $\phi= h'+u,$
($v,u$ denoting the vevs of the doublet and the singlet scalar respectively) 
and hence the corresponding squared mass matrix for scalars can be written as 
\begin{equation}\label{hmass}
 M^2_{\tilde{h},h'}=\left(
           \begin{array}{cc}
            2\lambda_1v^2  &   \lambda_{12}vu\\
            \lambda_{12}vu &   2\lambda_2u^2
                                    \end{array}
                                   \right).
\end{equation}

From (\ref{hmass}), we obtain the condition for having a stable potential 
\begin{eqnarray}
 \nonumber 4\lambda_1\lambda_2-\lambda_{12}^2&>&0,\\
\nonumber \lambda_{1,2}&>&0,  
\end{eqnarray}
known as co-positivity constraints~\cite{Kannike:2012pe} and the perturbativity 
constraints are given by $\lambda_{1},\lambda_{12}<4\pi$. Furthermore, in such 
scenarios (Eq. (\ref{eq:spot})), a detailed analysis concerning the vacuum 
stability of the potential can be obtained 
in~\cite{Lebedev:2012zw,EliasMiro:2012ay}. There will be two 
physical Higgses ($h$ and $H'$) whose 
 mass eigenvalues are given by 
\begin{eqnarray}
 \nonumber m_{h}^2&=&\lambda(v^2+u^2)-\sqrt{\lambda^2 (v^2-u^2)^2
            +\lambda_{12}^2u^2v^2}\label{mh1-2},\\
 m_{H'}^2&=&\lambda(v^2+u^2)+\sqrt{\lambda^2 (v^2-u^2)^2
            +\lambda_{12}^2u^2v^2}\label{mh2-2},
\label{mh2}
\end{eqnarray}
and their mixing is through
\begin{equation}
 \sin 2\theta = \frac{\lambda_{12}uv}{\sqrt{(\lambda_1v^2-\lambda_2u^2)^2
            +\lambda_{12}^2u^2v^2}}\label{cst-2}. 
\end{equation}

For simplicity, we consider $\lambda_1= \lambda_2=\lambda$ for 
our study and we will mostly follow this for the rest of the  paper. However, 
given a relaxation of this constraint, we will have one more parameter to 
control the DM phenomenology in particular. We have added a short analysis  
in appendix \ref{apnd2} mentioning possible modifications when $\lambda_1 \neq \lambda_2$ 
and its implication in DM phenomenology.

We have two mass eigenstates $h$ and $H'$ with mass eigenvalues  $m_h$ and
$m_{H'}$. Out of these two mass eigenstates, we can identify any one to be the 
observed Higgs boson discovered at LHC~\cite{Aad:2012tfa, Chatrchyan:2012xdj} 
with mass 125.7 GeV~\cite{Agashe:2014kda} depending on the choice of mixing. 
This can be done in two ways, namely, 

\begin{enumerate}
 \item {\it Low mass region:} Here we consider the additional scalar  
       to be lighter than the SM  Higgs discovered at LHC. Therefore, in this 
       case we write these mass eigenstates as $m_{H'} 
       =m_{ h_{ \rm SM }} $ and $m_h=m_{H_{\rm light}}$. Note that this is 
       viable, 
       as the other state is dominantly a singlet to avoid the collider search bounds. 
 \item {\it High mass region:}  Here we identify the additional scalar field to 
        be  heavier than the SM Higgs discovered at LHC. In this scenario we 
        consider the mass eigenstates to be $m_{h} 
       =m_{ h_{ \rm SM }} $ and $m_{H'}=m_{H_{\rm heavy}}$. 
\end{enumerate}
It is easy to understand that the mixing has two different limits 
for the above two cases to be phenomenologically viable. 
Following our notation, the decoupling limit corresponds to $\sin\theta\sim 1$ for the Low mass region and $\sin\theta\sim 0$ 
for the 
High mass region. We will address the two cases separately. Following ref \cite{Robens:2015gla, 
Robens:2016xkb}, it turns out that we have approximately $\sin\theta \gtrsim 0.9$  for low mass region ($ \lesssim 100$ GeV),
and $\sin\theta \lesssim 0.3$ for the high mass region ($\gtrsim 150$ GeV).
For demonstration purposes, we use values of $\sin\theta$ within these specified range in both cases,
without going to the details of this $\sin\theta$ dependence on the extra scalar mass and the ratio 
of the two vevs.

\subsection{Low mass region}

Following the notations introduced in earlier section, the relation between the mass and gauge eigenstates 
can be written as:
\begin{equation}\label{phi-low}
\left(
      \begin{array}{c}
                   H_{\rm light}\\
                  h_{\rm SM}
                   \end{array}
               \right)=\left(
                                \begin{array}{cc}
                      \cos\theta  &  -\sin\theta\\
                        \sin\theta &   \cos\theta
                                    \end{array}
                                   \right)\left(
      \begin{array}{c}
                   \tilde{h}\\
                  h'
                   \end{array}
               \right).
\end{equation}
Correspondingly, the masses in terms of the input parameters can be written as: 
\begin{eqnarray}
 \nonumber m_{H_{\rm light}}^2&=&\lambda_1v^2+\lambda_2 u^2-\sqrt{(\lambda_1 v^2-
                           \lambda_2 u^2)^2+\lambda_{12}^2u^2v^2}, \label{mH1}\\
 m_{h_{\rm SM}}^2  &=& \lambda_1v^2+\lambda_2 u^2+\sqrt{(\lambda_1 v^2-
                        \lambda_2 u^2)^2+\lambda_{12}^2u^2v^2}\label{mH2}. 
\end{eqnarray}
We recall that we identify $h_{\rm SM}$ to be the Higgs discovered at LHC (with 
$m_{h_{\rm SM}}=125.7$ GeV) and $m_{H_{\rm light}}<m_{h_{\rm SM}}$. Therefore, 
we need to choose large $\sin\theta$ limit ($\sin\theta\rightarrow 1$) 
for $h_{\rm SM}$ to get dominant contribution from the SM scalar doublet 
$H$ whereas $H_{\rm light}$ remains dominantly a scalar singlet 
(as can be seen from Eq. (\ref{phi-low})). From the above mentioned expressions, 
we can recast the couplings in the scalar 
potential in terms of the physical quantities like masses and the mixing angels
as
\begin{eqnarray}
 \lambda_1 &=& \frac{m_{H_{\rm light}}^2}{4v^2}(1+\cos 2\theta) +\label{l1low} 
              \frac{m_{h_{\rm SM}}^2}{4v^2}(1-\cos 2\theta)\label{l1},\\
 \lambda_2 &=& \frac{m_{H_{\rm light}}^2}{4u^2}(1-\cos 2\theta) + \label{l2low}
              \frac{m_{h_{\rm SM}}^2}{4u^2}(1+\cos 2\theta)\label{l2},\\
\lambda_{12}&=& \sin 2\theta \frac{m_{h_{\rm SM}}^2-
                 m_{H_{\rm light}}^2}{2uv}\label{l3}. 
\end{eqnarray}
Now with the consideration $\lambda_1= \lambda_2=\lambda$, 
we evaluate $\lambda$, $\lambda_{12}$ and $u$ for fixed values of $m_{H_{\rm 
light}}$ and mixing angle $\sin\theta$ using Eqs. (\ref{mH1})- (\ref{l2}). 
In Table \ref{tab:2} we evaluate the values of $u, \lambda, \lambda_{12}$, for 
two different choices of mixing angle $\sin\theta=0.999$ and 0.9 and three different choices 
of the light scalar mass $m_{H_{\rm light}}=60, ~80,~100$ GeV.

\begin{table}[h]
\centering
\begin{tabular}{|c|c|c|c|c|c|c|}
\hline
\multirow{2}{*}{\begin{tabular}{c}
 Benchmark Points\\
$m_{H_{\rm light}}$ (GeV)\\
\end{tabular}} & \multicolumn{2}{c|}{$u$ (GeV)} & 
\multicolumn{2}{c|}{$\lambda$} & \multicolumn{2}{c|}{$\lambda_{12}$}\\ 
\cline{2-7}
        &  $s_{\theta}=0.999$  &  $s_{\theta}=0.9$ &  $s_{\theta}=0.999$ &  
$s_{\theta}=0.9$ &  $s_{\theta}=0.999$ &  $s_{\theta}=0.9$ \\ \hline
    60      & 117.91 & 162.98 & 0.1303 & 0.1114 & 0.0188 & 0.1194
         \\ \hline
    80      & 156.89 & 188.01 & 0.1304 & 0.1158 & 0.0109 & 0.0797
         \\ \hline
   100    & 195.89 & 213.80 & 0.0827 & 0.1214 & 0.0034 & 0.0433
         \\ \hline
\end{tabular}
\caption{Vacuum Expectation value of singlet scalar ($u$) and the dimensionless couplings 
in the scalar potential ($\lambda_1=\lambda_2=\lambda, \lambda_{12}$) evaluated at 
some selected benchmark points for $m_{H_{\rm light}}$
=60, 80, and 100 GeV with $\sin\theta=0.999 (0.90), m_{h_{\rm SM}}=125.7$ GeV 
with SM Higgs vev $v=246$ GeV.
}
\label{tab:2}
\end{table}

\begin{figure}[h]
$$
\includegraphics[height=6cm]{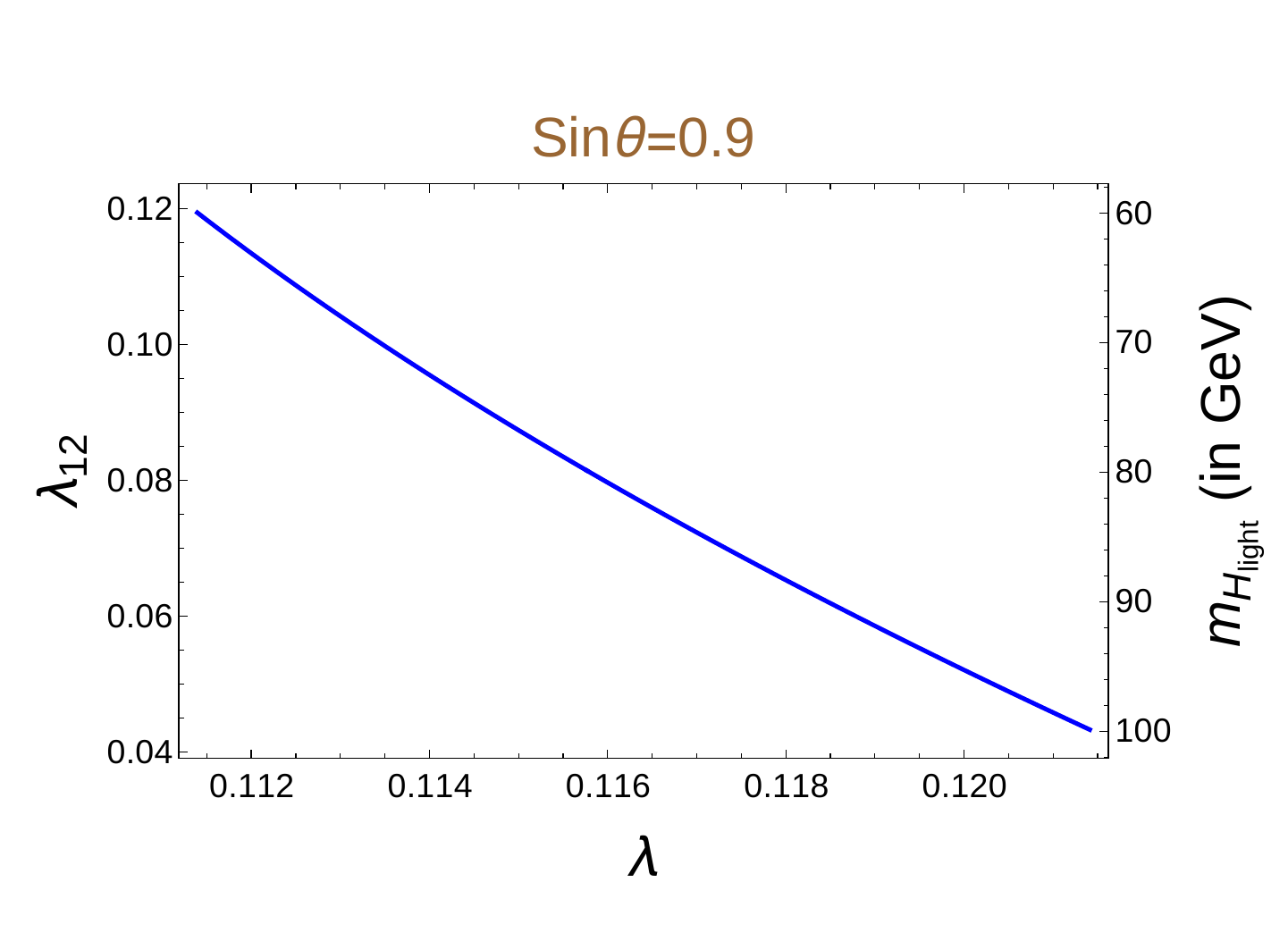}
\includegraphics[height=6cm]{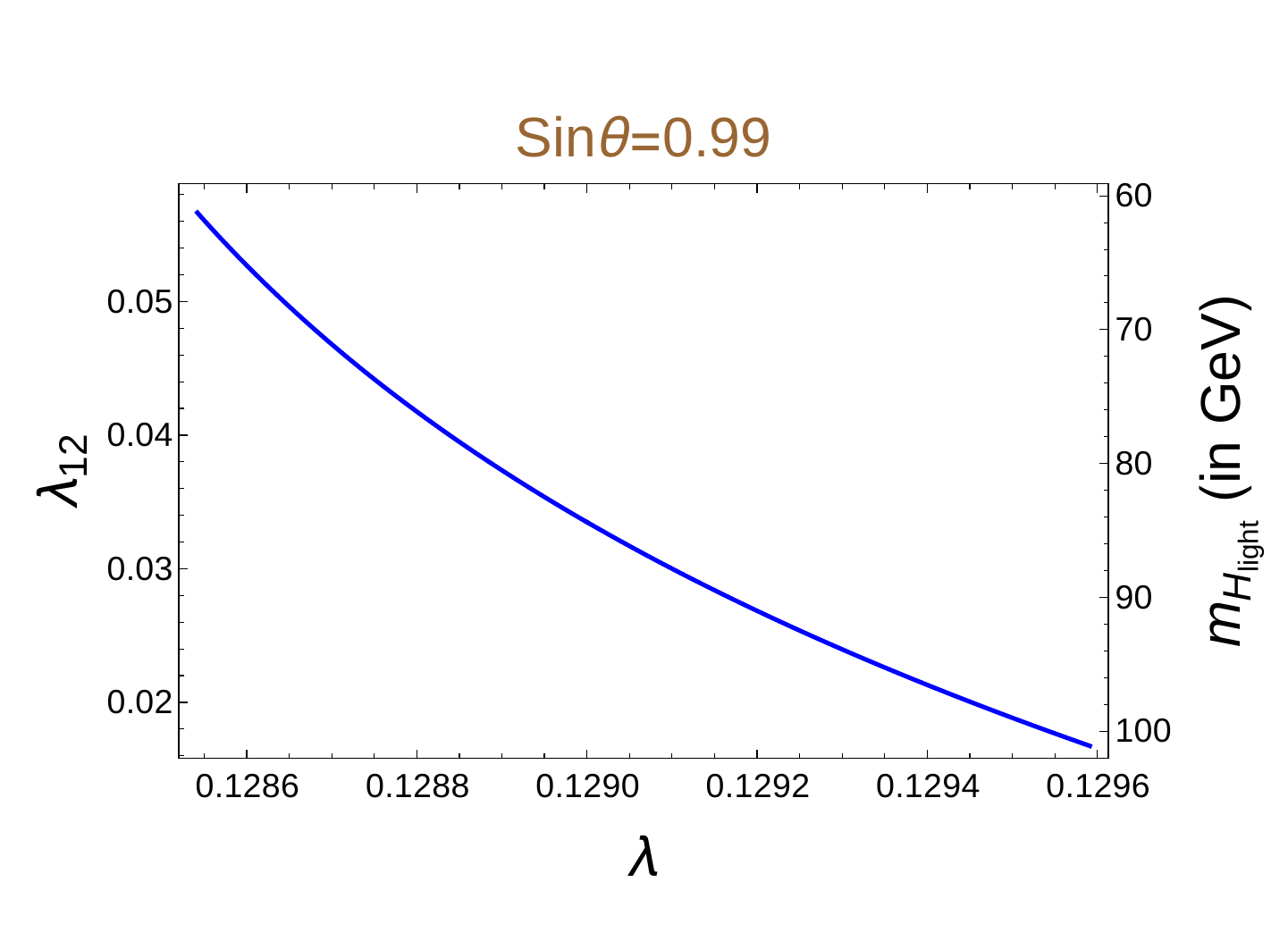}
$$
\caption{Correspondence between $\lambda=\lambda_1=\lambda_2$ and 
$\lambda_{12}$ for low mass region with the light scalar mass varying between 
60 GeV $\leq m_{H_{\rm light}} \leq $ 100 GeV, with mixing angle fixed at 
$\sin\theta=0.9$ (for left panel) and $\sin\theta=0.99$ (for right panel) 
respectively.}
\label{fig:mlow}
\end{figure}
In Fig \ref{fig:mlow} we show the variation of 
$\lambda$ and $\lambda_{12}$ for the low mass region. Here we observe that for 
a variation of the light Higgs mass $m_{H_{\rm light}}=60-100$ GeV, 
$\lambda$ varies between 0.11 to 0.12, whereas $\lambda_{12}$ varies 
between 0.12 to 0.04 when $\sin\theta$ is fixed at 0.9. Note here that there is a 
one-to-one correspondence between $\lambda$ and $\lambda_{12}$, 
which broadens up once we relax our assumption, i.e. $\lambda_1 \neq \lambda_2$ (see Appendix \ref{apnd2}). 
When $\lambda_{12}$ increases, $\lambda$ decreases. There is another point 
to note here, $\lambda_{12}$ is also present in the triple Higgs vertex, which 
will control the DM phenomenology to some extent as we will demonstrate. However,
we do not have a freedom of choosing it once we know both the Higgs masses and mixing. 
Self couplings are anyway very difficult to estimate in collider experiments. 
The analysis above is mainly aimed at showing 
the legitimacy of the input parameters for the choices of the masses of light Higgs, which 
we use in our further analysis.

\subsection{High mass region}


As in the low mass case, the relation between the mass eigenstates and 
gauge eigenstates for the high mass region can be written as,
\begin{equation}\label{phi-high}
\left(
      \begin{array}{c}
                  h_{\rm SM} \\
                  H_{\rm heavy}
                   \end{array}
               \right)=\left(
                                \begin{array}{cc}
                      \cos\theta  &  -\sin\theta\\
                        \sin\theta &   \cos\theta
                                    \end{array}
                                   \right)\left(
      \begin{array}{c}
                   \tilde{h}\\
                  h'
                   \end{array}
               \right). 
\end{equation}
with the corresponding physical masses given by
\begin{eqnarray}
 \nonumber m_{h_{\rm SM}}^2&=&\lambda_1v^2+\lambda_2 u^2-\sqrt{(\lambda_1 v^2
                   -\lambda_2 u^2)^2+\lambda_{12}^2u^2v^2},\\
 m_{H_{\rm heavy}}^2&=& \lambda_1v^2+\lambda_2 u^2+\sqrt{(\lambda_1 v^2
                    -\lambda_2 u^2)^2+\lambda_{12}^2u^2v^2}.
\end{eqnarray}

\begin{figure}[h]
$$
\includegraphics[height=6cm]{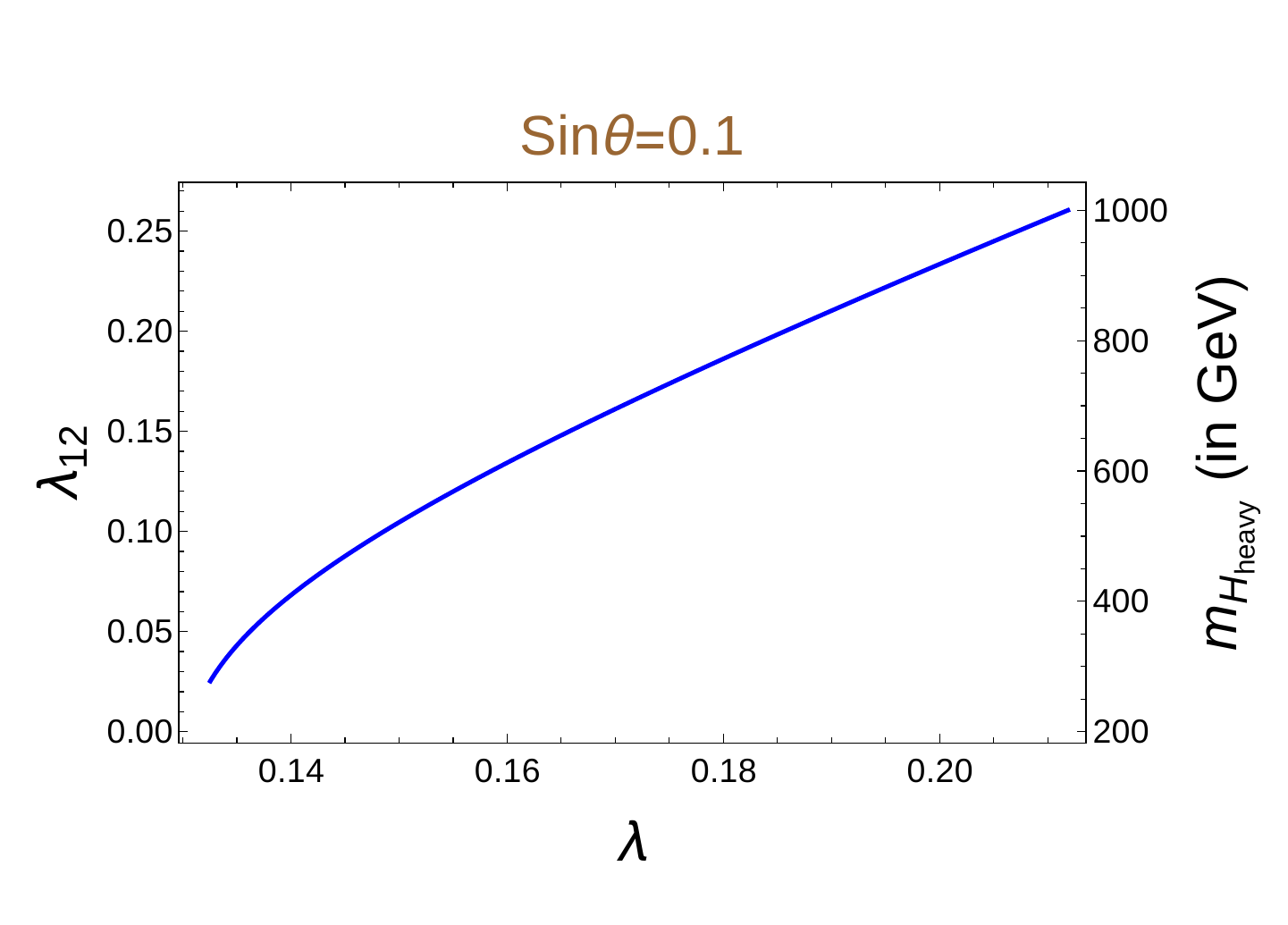}
\includegraphics[height=6cm]{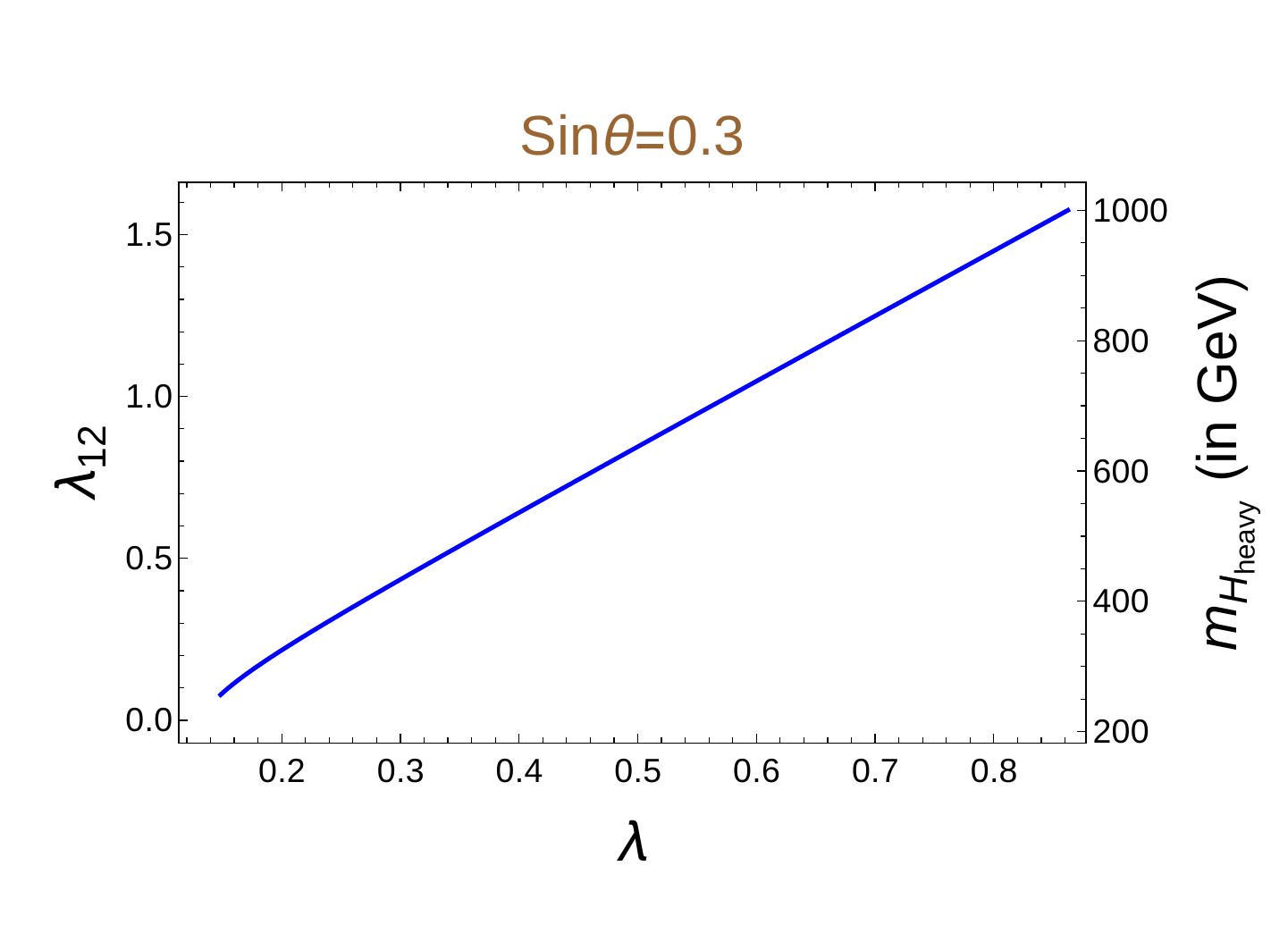}
$$
\caption{The correlation of $\lambda$ vs $\lambda_{12}$ for high mass region
with heavy Higgs ranging between 200 GeV $\leq m_{H_{\rm heavy}} \leq $ 1000 
GeV for $\sin\theta=0.1$ (left panel) and  $\sin\theta=0.3$ (right panel) 
respectively.}
\label{fig:mhigh}
\end{figure}

Following the notation we established, we identify $h_{\rm SM}$ as the Higgs discovered at the LHC (with 
$m_{h_{\rm SM}}=125.7$ GeV) and $m_{h_{\rm SM}}< m_{H_{\rm heavy}}$. 
Therefore, in this alternate scenario, we work in small $\sin\theta$ limit 
($\sin\theta\rightarrow 0$ for decoupling), where $h_{\rm SM}$ is dominantly 
a scalar doublet and $H_{\rm heavy}$ gets contribution mainly from the scalar singlet 
$\phi$. This gives us the freedom to choose any mass of the other physical scalar
heavier than the observed Higgs mass. To see how the physical masses are related to the input parameters,
we recast the couplings as:
\begin{eqnarray}
 \lambda_1 &=& \frac{m_{h_{\rm SM}}^2}{4v^2}(1+\cos 2\theta)\label{l1high}
            + \frac{m_{H_{\rm heavy}}^2}{4v^2}(1-\cos 2\theta),\\
 \lambda_2 &=& \frac{m_{h_{\rm 
SM}}^2}{4u^2}(1-\cos 2\theta)\label{l2high}
            + \frac{m_{H_{\rm heavy}}^2}{4u^2}(1+\cos 2\theta),\\
\lambda_{12}&=& \sin 2\theta \frac{m_{H_{\rm heavy}}^2-m_{h_{\rm SM}}^2}{2uv}.
\end{eqnarray}
In order to estimate the couplings in the high mass region, we choose $\sin\theta=0.3$ and 0.001 (values of the mixing angle
near the two extremes that are admissible by Higgs data for this region). Using these mixing angles, we find
$u, \lambda~(=\lambda_1=\lambda_2)$ and $\lambda_{12}$ for various values of $m_{H_{\rm heavy}}$ in Table 3. For larger 
masses and larger mixing angle, the 
couplings $\lambda$ and $\lambda_{12}$ are also larger. 
In Fig. \ref{fig:mhigh}, we have presented the correlation between 
$\lambda$ and $\lambda_{12}$ for the high mass region, fixing $\sin\theta$ at 
a moderate value of 0.1. For the heavy Higgs mass ranging between 
$m_{H_{\rm heavy}}$ = 200-1000 GeV, we find that $\lambda$ and $\lambda_{12}$ 
varies between 0.13-0.21 and 0.025-0.26 respectively. For this case, we also see that for larger
$\lambda$, $\lambda_{12}$ is also larger. Again, if we relax the 
condition of $\lambda_1=\lambda_2$, the correspondence will be relaxed.


\begin{table}[h]
\centering
\begin{tabular}{|c|c|c|c|c|c|c|}
\hline
\multirow{2}{*}{\begin{tabular}{c}
 Benchmark Points\\
$m_{H_{\rm heavy}}$ (GeV)\\
\end{tabular}} & \multicolumn{2}{c|}{$u$ (GeV)} & 
\multicolumn{2}{c|}{$\lambda$} & \multicolumn{2}{c|}{$\lambda_{12}$}\\ 
\cline{2-7}
        &  $s_{\theta}=0.3$  &  $s_{\theta}=0.001$ &  $s_{\theta}=0.3$ &  
$s_{\theta}=0.001$ &  $s_{\theta}=0.3$ &  $s_{\theta}=0.001$ \\ \hline
    200  & 356.81 & 391.41 & 0.1485 & 0.1306 & 0.0789 & 0.0003
         \\ \hline
    400 & 556.02 & 782.81 & 0.2378 & 0.1306 & 0.3017 & 0.0008
         \\ \hline
    600 & 662.42 & 1174.21 & 0.3865 & 0.1306 & 0.6138 & 0.0012
         \\ \hline
    800 & 700.61 & 1565.60 & 0.5947 & 0.1306 & 1.0365 & 0.0016
         \\ \hline
   1000 & 726.93 & 1956.98 & 0.8624 & 0.1306 & 1.5751 & 0.0020
         \\ \hline
\end{tabular}
\caption{Values obtained for $\lambda_1=\lambda_2=\lambda$ and vev $u$ (in GeV)
for some chosen heavy Higgs masses $m_{H_{\rm heavy}}$
= 200, 400, 600 and 1000 GeV and two extreme values of mixing angles 
$\sin\theta=0.3~(0.001), m_{h_{\rm SM}}=125.7$ GeV and $v=246$ GeV. 
}
\label{tab:3}
\end{table}

\section{Dark matter relic density and direct search constraints}\label{sec:dm}

In this framework the dark sector is kept minimal and consists of a SM singlet 
fermion $\chi$. A bare Majorana mass term for this fermion is forbidden due to the specific 
$Z_4$ charge as assumed and shown in Table \ref{tab:1}. The presence of $Z_4$ 
charged scalar field $\phi$, yields the only coupling  $y \phi \bar{\chi^c} 
\chi$, involving the dark fermion as shown in the Lagrangian (Eq. (\ref{lag})). 
The mass of dark matter ($M_{\chi}$) appears only through this interaction term, once $\phi$ acquires 
a vev. At this point, we note that there is no 
direct renormalisable interaction of the DM with SM fields, which 
is true for any singlet fermion DM, unless extended to the presence of an 
additional doublet~\cite{ArkaniHamed:2005yv, Mahbubani:2005pt, DEramo:2007anh, 
Enberg:2007rp, Cynolter:2008ea, Cohen:2011ec, Cheung:2013dua, 
Restrepo:2015ura, Calibbi:2015nha, Freitas:2015hsa, 
Bizot:2015zaa, Cynolter:2015sua}. As we have already mentioned, the scalar 
field ($\phi$), which connects to the dark sector, also plays a crucial role in the 
neutrino mass generation and the explicit connection between the dark and 
neutrino sectors will be demonstrated shortly.  Thus, prior to the spontaneous breaking of this 
$Z_{4}$ symmetry, the DM remains massless ($M_{\chi}=0$) and doesn't 
have any connection to the visible sector. After SSB, $\phi$ acquires a vev 
$\langle \phi \rangle$, mixes with SM Higgs doublet ($H$) through the $\phi^2 
H^\dagger H$ term and connects the DM with the SM. The two phenomenologically 
viable scenarios discussed above after SSB, with the additional physical scalar having either lower or higher mass  
than the physical scalar seen at the LHC, will lead to 
the following DM couplings through the interactions (following earlier 
notation): 

1. Low mass region (large $\sin\theta$):
\begin{eqnarray}\label{dmc:low}
 y \phi \bar{\chi^c} \chi& =&y(-H_{\rm light}\sin\theta+h_{\rm SM}\cos\theta) 
         \bar{\chi^c} \chi ,\nonumber\\
 &= &\frac{M_{\chi}}{u} (-H_{\rm light}\sin\theta+h_{\rm SM}\cos\theta) 
\bar{\chi^c} \chi
\end{eqnarray}

2. High mass region (small $\sin\theta$):
\begin{eqnarray}\label{dmc:high}
  y \phi \bar{\chi^c} \chi& = &y(-h_{\rm SM}\sin\theta+H_{\rm heavy}\cos\theta) 
\bar{\chi^c} \chi,\nonumber\\
    &=& \frac{M_{\chi}}{u} (-h_{\rm SM}\sin\theta+H_{\rm heavy}\cos\theta)\bar{\chi^c} \chi.
\end{eqnarray}
\begin{figure}[h]
$$
\includegraphics[height=4cm]{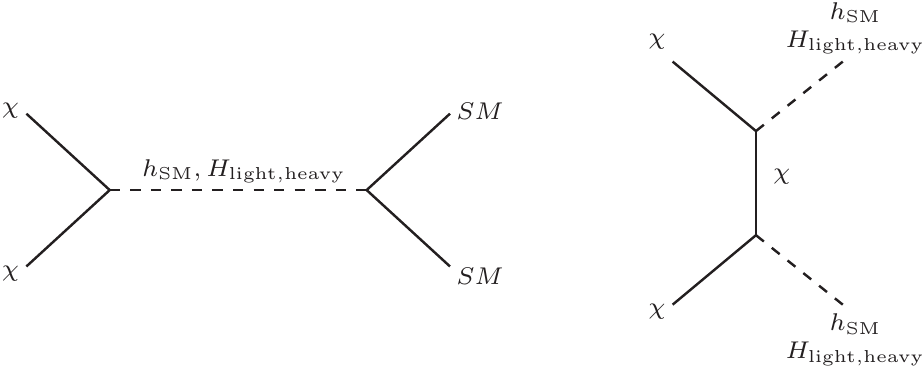}
$$
\caption{DM ($\chi$) annihilation process that yield relic density of the DM.}
\label{fig:1}
\end{figure}
In Eq. (\ref{dmc:low}) and (\ref{dmc:high}), we have replaced the DM-scalar coupling 
by the DM mass as: $y=M_{\chi}/ \langle \phi \rangle =M_{\chi}/u$. 
Note here, we will adhere to the conservative limit of $y \le 1.0$ from perturbative 
constraints (see Appendix A) assuming the cut-off scale of the theory to be around $\Lambda \sim 10^7$ GeV. 
The limit on $\Lambda$ in this analysis is obtained satisfying neutrino mass constraints and will be detailed in
Sec. \ref{sec:conec}.
The DM will then have an s-channel annihilation through 
the two physical Higgses to the SM. The other possibility will be to have a 
t-channel annihilation to the Higgses as shown in the Feynman graphs in Fig. \ref{fig:1}. 
These processes will help the fermion DM to thermally freeze out and we will compute 
the required values of the DM parameters to satisfy relic density constraints.
The second processes (t-channel graphs) will only be feasible when the DM mass is 
heavier than the Higgs masses. The t-channel graphs will also play an important role to 
disentangle the annihilation processes to the possible direct search cross-sections that the DM 
obtains. With non-observation of DM in direct search experiments, such processes are essential
to keep the DM model viable. 

Here we again mention that the DM analysis has been performed considering 
$\lambda$ $=\lambda_1=\lambda_2$  and $m_{h_{\rm SM}}$ fixed at 125.7 GeV 
as observed by the LHC. Subsequently, we can evaluate $u, 
\lambda$ and $\lambda_{12}$ once $\sin\theta$ and $m_{H_{\rm light,~heavy}}$ 
are known. Choice of these two parameters ($\sin\theta$ and $m_{H_{\rm 
light,~heavy}}$) is constrained by perturbative unitarity, 
EW precision data, perturbativity of the couplings along with vacuum stability. 
We refrain from a detailed discussion in this regard, which can be found in 
\cite{Robens:2015gla, Robens:2016xkb}. 

Hence the DM phenomenology is completely dictated by three parameters: 
\begin{equation}
 M_{\chi}, \sin\theta {~\rm and~} m_{H_{\rm light,~heavy}}  \nonumber
\end{equation}
where  $M_{\chi}$ is the mass of the DM,  $\sin\theta$ is the mixing 
angle between two scalars and $m_{H_{\rm light,~heavy}}$ is the mass of the 
additional scalar field. In our analysis, we use  
micrOMEGAs 4.2.5 \cite{Belanger:2008sj} to find the relic density and direct search 
cross sections by scanning the three-fold DM parameter space within the admissible limits. 

Depending upon the mass of $m_{H_{\rm light,~heavy}}$, our analysis 
is categorised in two different scenarios: Low mass region ($m_{H_{\rm 
light}}<m_{h_{\rm SM}}$) and High mass region ($m_{h_{\rm SM}}< m_{H_{\rm 
heavy}}$) respectively to obtain constraints from DM relic density 
and non-observation of DM in direct search experiments. 

Before we go into the details of the parameter space scan and the associated DM constraints, 
a few comments are in order. The vev of $\phi$, as is clearly seen from the
Lagrangian, will break the $Z_4$ symmetry to a remnant $Z_2$. The stability of 
the DM is ensured by this preserved $Z_2$ symmetry. The situation of a SM singlet fermion DM that is $Z_2$-odd and
connected to the SM through the mixing of a scalar singlet (which is even under $Z_2$) with
SM Higgs has been studied in the literature~\cite{Kim:2008pp, 
Baek:2011aa, LopezHonorez:2012kv, Carpenter:2013xra, Abdallah:2015ter, 
Dutra:2015vca, Buckley:2014fba, DeSimone:2016fbz, Arcadi:2017kky}. The model we 
consider here has distinct DM phenomenology, although 
mostly having similar features. One of the main differences is simply that the 
dark matter mass is proportional to the vev of the extra scalar $\phi$ (which in turn is related 
to $\sin\theta$). Another is the absence of the cubic $\phi^3$ term in the scalar 
potential, forbidden by the $Z_4$ symmetry in our model, but allowed in the previously studied models where the 
scalar $\phi$ would be neutral. The cubic term would have altered 
the triple Higgs vertex, therefore changing the 
s-channel annihilation for the DM to the Higgs final states and also adding to the freedom of choosing the Higgs masses 
while keeping the input parameters within admissible range.
The $Z_4$ symmetry in our model that is essential to connect in our model the DM to the 
neutrino sector, is therefore further motivated as it simplifies the DM analysis, and makes the model considerably more
predictive.
The scans performed 
in the next subsections have been systematized to the requirement of connecting to the neutrino sector.

\subsection{Dark matter phenomenology in light Higgs mass region}

As has already been mentioned, in this case we assume the second neutral Higgs to be lighter 
$m_{h_{\rm SM}}>m_{H_{\rm light}}$ with large $\sin\theta$ and of course
$m_{h_{\rm SM}}=125.7$ GeV. In Table \ref{tab:4} we list relevant vertices that 
connects the DM to the visible sector and corresponding vertex factors in terms of parameters
$M_{\chi}, \sin\theta$ and vevs $u,v$ and couplings $\lambda, \lambda_{12}$ with the assumption of
$\lambda_1=\lambda_2=\lambda$. Note that the couplings are only present in the triple Higgs vertex
and they do not show up elsewhere.  Also we note again that the the couplings are automatically determined
once we choose the light scalar mass and mixing. The vertices are introduced in the code 
micrOMEGAs~\cite{Belanger:2008sj} for a 
scan of the parameters to yield correct relic density and direct search 
observations.  

\begin{table}[h]
\centering
\label{my-label}
\begin{tabular}{|c|l|}
\hline
Vertices & Vertex Factor    \\
\hline
\hline 
$H_{\rm light}\chi\chi$ &  $-y\sin\theta=-(M_{\chi}/u)\sin\theta$\\
\hline
$H_{\rm light}W^+W^-$   &  $\frac{2M_W^2}{v}\cos\theta$\\
\hline
$H_{\rm light} Z Z$   &  $\frac{2M_Z^2}{v}\cos\theta$\\
\hline
$H_{\rm light} h_{\rm SM} h_{\rm SM}$   &  
$2\left(\frac{1}{2}v\lambda_{12}c^3_{\theta}-3u\lambda 
c^2_{\theta}s_{\theta}+u\lambda_{12}
c^2_{\theta}s_{\theta}+3v\lambda c_{\theta}s^2_{\theta}- v 
\lambda_{12}c_{\theta}s^2_{\theta}-\frac{1}{2}u\lambda_{12}s^3_{\theta}
\right)$\\

\hline
\hline
$h_{\rm SM}\chi\chi$ &  $y\cos\theta=(M_{\chi}/u)\cos\theta$\\
\hline
$h_{\rm SM}W^+W^-$   &  $\frac{2M_W^2}{v}\sin\theta$\\
\hline
$h_{\rm SM} Z Z$   &  $\frac{2M_Z^2}{v}\sin\theta$\\
\hline
$h_{\rm SM} f \overline{f}$   &  $\frac{m_f}{v}\sin\theta$\\
\hline
$h_{\rm SM} H_{\rm light} H_{\rm light}$   &  
$2\left(\frac{1}{2}u\lambda_{12}c^3_{\theta}+3v\lambda 
c^2_{\theta}s_{\theta}-v\lambda_{12}
c^2_{\theta}s_{\theta}+3u\lambda c_{\theta}s^2_{\theta}- u 
\lambda_{12}c_{\theta}s^2_{\theta}+\frac{1}{2}v\lambda_{12}s^3_{\theta}
\right)$\\
\hline
\end{tabular}
\caption{Vertices that connect DM to the SM in low mass region. Vertex factors are written assuming 
$\lambda_1=\lambda_2=\lambda$.}\label{tab:4}
\end{table}

\subsubsection{Relic Density}

\begin{figure}[h]
$$
\includegraphics[height=5.2cm]{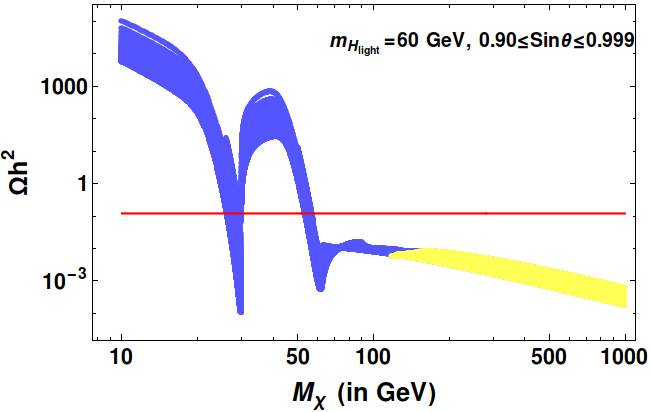}
\includegraphics[height=5.2cm]{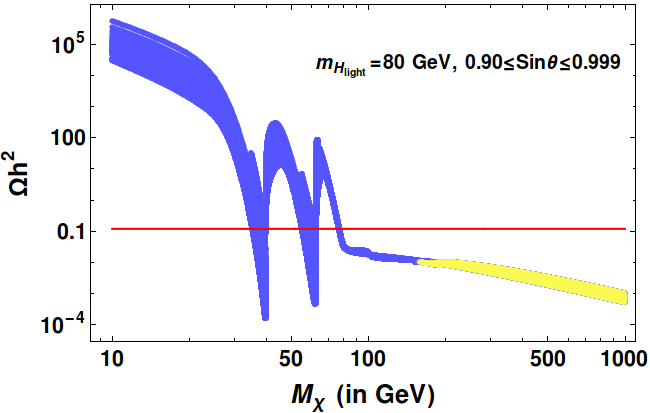}
$$
$$
\includegraphics[height=5.2cm]{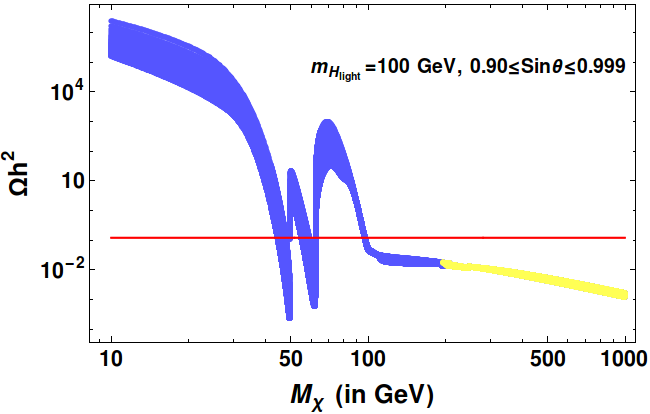}
$$
\caption{Relic density ($\Omega h^2$) of the fermion DM as a function of DM mass 
 $M_{\chi}$ for three different choices of $m_{H_{\rm light}}$=60, 80 and 100 
GeV respectively shown in the upper left, right and bottom panel. The Higgs mixing angle 
is varied between $0.9 \leq \sin\theta \leq 0.99$. Yellow region lies outside perturbative limit of coupling $y$ 
and corresponds to $y>1.0$. Horizontal red line represents observed relic density by Planck data (see Eq. (\ref{relic-limit})).}
\label{fig:omega-1ow}
\end{figure}

The relic density of the DM is inversely proportional to the thermal averaged annihilation 
cross-section of the DM ($\langle \sigma {\rm v}\rangle$) guided by the relation (assuming $x_f \sim 20$):
\begin{equation}
\Omega h^2  \simeq \frac{2.4 \times 10^{-10}~{\rm GeV^{-2}}}{\langle \sigma \; {\rm v}\rangle_{_{\chi \chi \to SM \;SM}}},
\label{eq:Omega-sigmav}
\end{equation}
where $\langle \sigma \; {\rm v}\rangle_{_{\chi \chi \to SM \;SM}}$ denotes thermal average annihilation cross-section of the 
DM ($\chi$) to SM final states including the  light/heavy Higgs whenever kinematically permissible (as in Feynman graphs in Fig. \ref{fig:1}).
      Note that we obtain the relic density numerically by implementing the model in the code 
micrOMEGAs~\cite{Belanger:2008sj}.

In Fig. \ref{fig:omega-1ow}, we show the variation of DM relic density as a function of 
DM mass $M_{\chi}$. In the left, middle and right panel of Fig. \ref{fig:omega-1ow}, 
the BSM scalar mass $m_{H_{\rm light}}$ is fixed at 60, 80 and 100 GeV respectively. 
In each panel the blue patch represents the variation of the mixing angle $\sin\theta$ in the range 
$0.90\leq\sin\theta \leq 0.999$~\cite{Robens:2015gla, Robens:2016xkb}.
The region between the red horizontal lines represents correct relic density 
satisfying the PLANCK constraint~\cite{Ade:2015xua}:
\begin{equation}\label{relic-limit}
0.1175\leq \Omega h^2\leq 0.1219.
\end{equation}
Due to s-channel annihilation of the DM through the two Higgses ($h_{\rm SM}$ 
and $H_{\rm light}$), in each panel of Fig. \ref{fig:omega-1ow}, we find two resonance regions for 
$M_\chi=m_{h_{\rm SM}}/2$ and $M_\chi=m_{H_{\rm light}}/2$, where the relic density drops sharply, intersecting the relic density constraint. In these plots we also 
observe that, as soon as the dark matter mass becomes comparable (or greater than) with 
the mass of additional scalar $H_{\rm light}$, the dark matter annihilation is 
dominantly controlled by the $\chi\chi \rightarrow {H_{\rm light}} {H_{\rm light}}$ process through t-
channel graph (see Feynman diagram in Fig. \ref{fig:1}). In Fig. 
\ref{fig:omega-1ow}, we also find that the relic density approaches the constraint when $M_{\chi}\sim m_{H_{\rm light}}$ = 80 and 100 
GeV as shown in the right and bottom panel respectively. However for $m_{H_{\rm 
light}}=60$ GeV, due to large annihilation via $\chi\chi \rightarrow {H_{\rm 
light}} {H_{\rm light}}$, the only regions which satisfy 
observed relic density are the resonance regions. Very importantly, we note that the 
heavy dark matter mass regions are severely constrained by the perturbative limit  
on coupling $y$; imposing $y>1.0$ (regions marked in yellow), DM masses above 
$\sim 200$ GeV for light Higgs mass $\sim 100$ GeV are disfavoured. The perturbative limit is even 
stronger for smaller values of the light Higgs mass.
We also note that the DM mass $< m_{h_{\rm SM}}/2$ is constrained by the Higgs invisible 
decay branching fraction that limits the mixing angle $\sin\theta$, as detailed in Appendix \ref{apnd3}.

\begin{figure}[h!]
$$
\includegraphics[height=5cm]{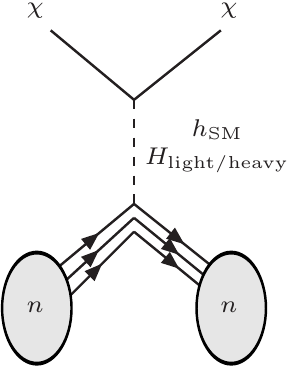}
$$
\caption{Feynman diagrams for DM to interact with Nucleon.}
\label{fig:dd}
\end{figure}

\subsubsection{Direct Search}
Direct search of the fermion DM occurs through the t-channel graph mediated by the Higgs
portal interactions as shown in Fig. \ref{fig:dd}. We again compute the direct 
search cross-sections of the DM through the code micrOMEGAs. 
The spin-independent Higgs portal interaction includes not just interactions with the light quarks, but also a dominant Higgs-gluon-gluon effective interaction \cite{Hisano:2010ct}. This gluon contribution arises at loop level, mediated by heavy quarks, and it is implemented in the code by the 
gluon form factor $f^{(n)}_{T_g}=\frac{2}{27}(1-\sum\limits_{q=u,d,s}f^{(n)}_{T_q})$ \footnote {The code also 
computes the QCD corrections to the effective Gluon form factor in the presence of any additional coloured particles,
in cases where the model contains them (e.g. in Supersymmetry).} \cite{Belanger:2008sj}. We use the default form factors of micrOMEGAs with $f^p_{T_u}=0.0153$, 
$f^p_{T_d}=0.0191$ and $f^p_{T_s}=0.0447$ for proton.
\begin{figure}[h]
$$
\includegraphics[height=5.2cm]{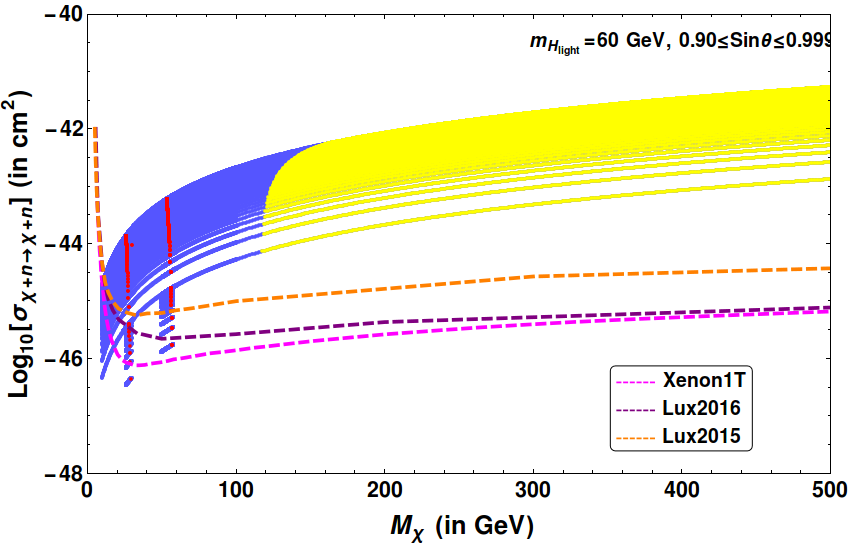}
\includegraphics[height=5.2cm]{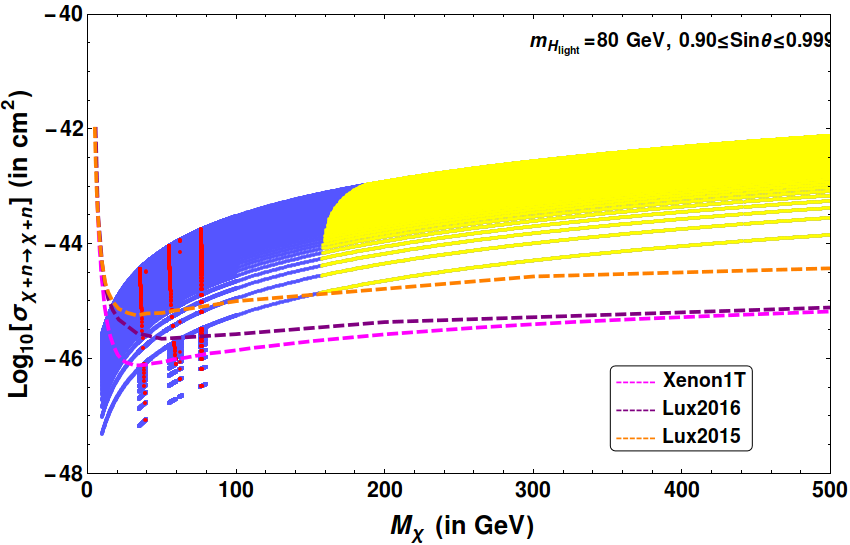}
$$
$$
\includegraphics[height=5.2cm]{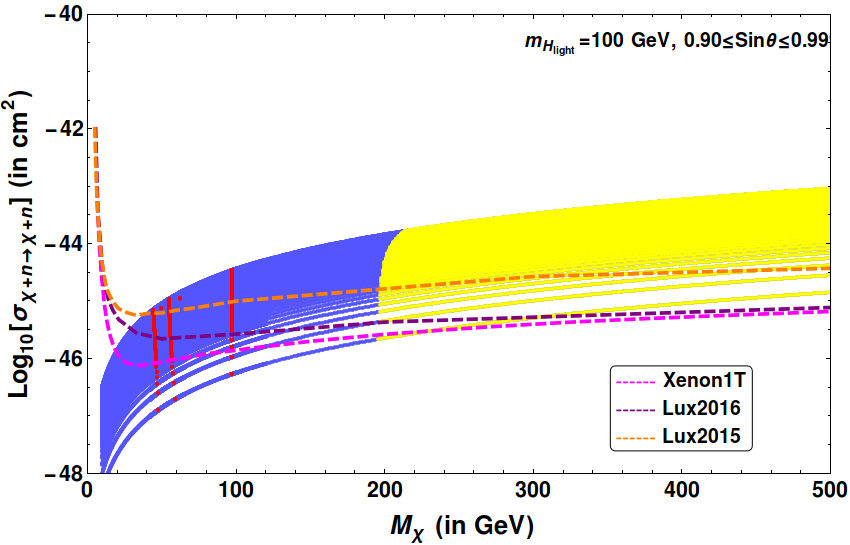}
$$
\caption{Spin independent direct
search cross section as a function of DM mass $M_{\chi}$ for three different 
choices of $m_{H_{\rm light}}$ =  60, 80 and 100 GeV respectively shown in the 
upper left, right and bottom panel. In each panel blue patch stands for 
$0.90\leq\sin\theta\leq 0.999$. Red dots in respective panels satisfy relic 
density constraints. Bounds from LUX and XENON1T are also shown through
dotted/dashed lines. The yellow region corresponds to $y>1.0$ which may violate the perturbative limit.}
\label{fig:direct-low}
\end{figure}
In Fig. \ref{fig:direct-low}, we show spin direct search scattering cross-section 
of the fermion DM in the low mass region. The plot is drawn as a function of DM mass 
$M_{\chi}$; for three distinct choices of the BSM scalar mass: 
$m_{h_{\rm light}}$ = 60 GeV (left), 80 GeV (right) and 100 GeV (bottom) respectively. 
We consider here the Higgs mixing $\sin\theta$ between 0.90 to 0.999, as we considered previously. 
The relic density allowed parameter space for the DM in these plots is represented by the dotted red dotted lines appearing in the blue regions. The current experimental bounds 
from non-observation of DM in direct search experiments from LUX and XENON1T data are shown by 
dashed lines. 

The main outcome of this analysis is to see that the regions which satisfy both 
relic density and direct search constraints are those in the resonance regions, 
$M_\chi=m_{h_{\rm SM}}/2, m_{H_{\rm light}}/2$
and  $M_{\chi}=m_{H_{\rm light}}$. The model is also disfavoured for DM mass beyond $\sim$ 200 GeV, due to the imposition of $y>1.0$ associated with the perturbative limit (see Appendix A).
Therefore, the model is quite restrictive in predicting the DM mass from DM 
constraints. This serves as a key feature to identify the connection 
to the neutrino sector in an unambiguous way.

\subsection{Dark matter phenomenology in heavy Higgs mass region}

Now let us turn to DM phenomenology of the heavy Higgs mass region, where we consider 
$m_{h_{\rm SM}}<m_{H_{\rm heavy}}$ with small $\sin\theta$ and $m_{h_{\rm SM}}=125.7$ GeV.  
First we note the vertices relevant for DM annihilations and scattering in Table 
\ref{tab:5} with appropriate vertex factors. This table is similar to Table 
\ref{tab:4} that corresponds to the low mass region excepting for the flip of notation 
in the triple Higgs vertices. Again, we parametrise the vertex factors in the limit of 
$\lambda_1=\lambda_2=\lambda$, which automatically get determined by the input of the
heavy Higgs mass and mixing. 

\begin{table}[h]
\centering
\label{my-label}
\begin{tabular}{|c|l|}
\hline
Vertices & Vertex Factor    \\
\hline
\hline 

$h_{\rm SM}\chi\chi$ &  $-y\sin\theta=-(M_{\chi}/u)\sin\theta$\\
\hline
$h_{\rm SM}W^+W^-$   &  $\frac{2M_W^2}{v}\cos\theta$\\
\hline
$h_{\rm SM} Z Z$   &  $\frac{2M_Z^2}{v}\cos\theta$\\
\hline
$h_{\rm SM} f \overline{f}$   &  $\frac{m_f}{v}\cos\theta$\\
\hline
$h_{\rm SM} H_{\rm heavy} H_{\rm heavy}$   &  $2\left(\frac{1}{2}v\lambda_{12} 
c^3_{\theta}-3u\lambda c^2_{\theta}s_{\theta} +u\lambda_{12} c^2_{\theta} 
s_{\theta} + 3v\lambda c_{\theta}s^2_{\theta}- v \lambda_{12} c_{\theta} 
s^2_{\theta} -\frac{1}{2} u \lambda_{12} s^3_{\theta}\right)$

\\
\hline

$H_{\rm heavy}\chi\chi$ &  $y\cos\theta=(M_{\chi}/u)\cos\theta$\\
\hline
$H_{\rm heavy}W^+W^-$   &  $\frac{2M_W^2}{v}\cos\theta$\\
\hline
$H_{\rm heavy} Z Z$   &  $\frac{2M_Z^2}{v}\cos\theta$\\
\hline
$H_{\rm heavy} h_{\rm SM} h_{\rm SM}$   &  $2\left(\frac{1}{2}u\lambda_{12} 
c^3_{\theta}+3v\lambda c^2_{\theta}s_{\theta} - v\lambda_{12} c^2_{\theta} 
s_{\theta}+ 3u\lambda c_{\theta} s^2_{\theta}- u \lambda_{12} c_{\theta} 
s^2_{\theta}+\frac{1}{2}v\lambda_{12} s^3_{\theta}\right)$\\
\hline
\end{tabular}
\caption{Relevant vertices implying DM-SM interactions with vertex factors 
for high mass region.}\label{tab:5}
\end{table}

\subsubsection{Relic Density}

\begin{figure}[h]
$$
\includegraphics[height=5.2cm]{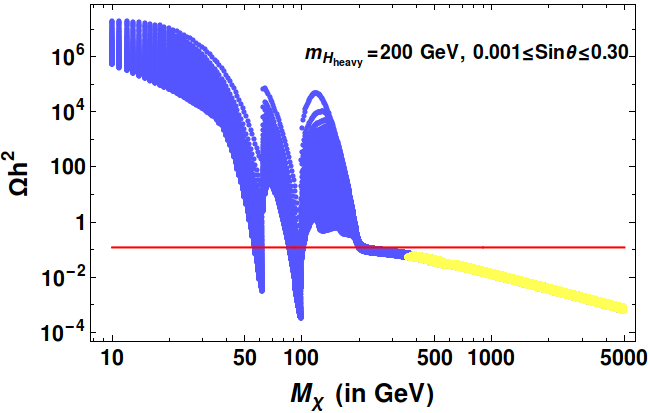}
\includegraphics[height=5.2cm]{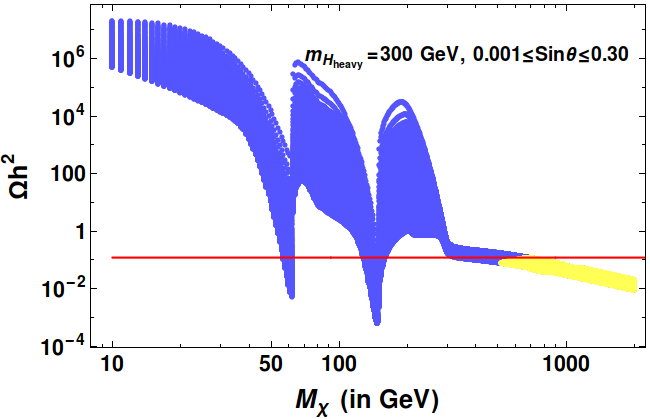}
$$
$$
\includegraphics[height=5.2cm]{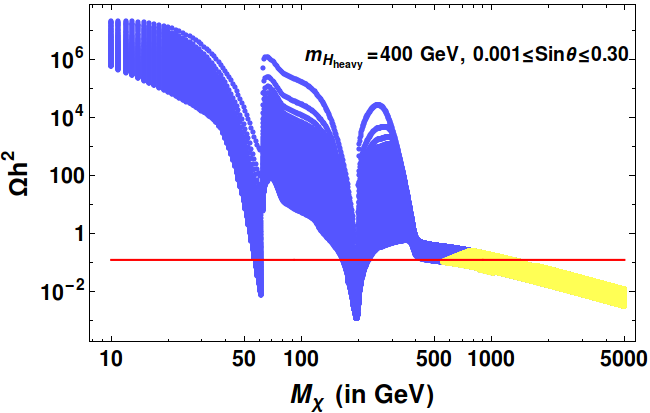}
\includegraphics[height=5.2cm]{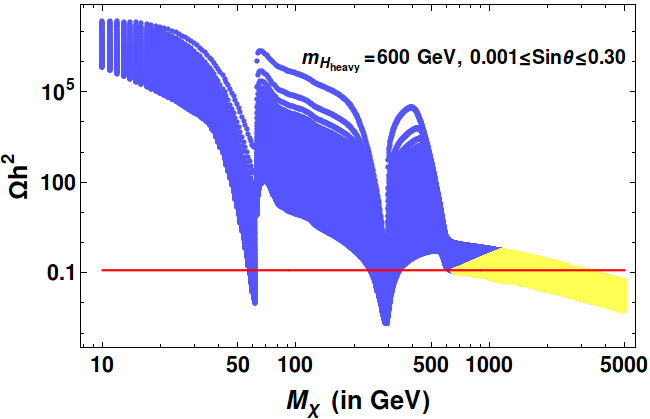}
$$
\caption{Relic density  ($\Omega h^2$) obtained as a function of DM mass ($M_{\chi}$) for three different 
choices of heavy Higgs mass: $m_{H_{\rm Heavy}}$=200, 300, 400 and 600 GeV. Mixing angle is 
varied between $0.001\leq\sin\theta \leq 0.3$. Horizontal red band represents 
observed relic density by Planck data. The yellow region corresponds to $y>1.0$ which may violate the perturbative limit.
}
\label{fig:omega-high}
\end{figure}

In this case relic density is plotted in Fig. \ref{fig:omega-high} for three 
different values for $m_{H_{\rm heavy}}$, namely 200, 300, 400, and 600 
GeV respectively. The mixing angle $\sin\theta$ is varied from
0.001 to 0.3~\cite{Robens:2015gla, Robens:2016xkb}. Here also, in each figure,  
two distinct resonances can be observed at $m_h/2$ and $m_{H_{\rm heavy}}/2$.
Apart from these resonance regions, for each choice of $m_{H_{\rm heavy}}$, 
there exists a large allowed range for $M_{\chi}>m_{H_{\rm heavy}}$, 
which satisfies the observed relic density by Planck data, mainly dominated by the annihilation channel 
$\chi\chi \rightarrow {H_{\rm heavy}} {H_{\rm heavy}}$. In the 
high mass region there is a relative large span of the mixing angle (when compared to the low mass region). Another important point is
that the triple Higgs vertex is proportional to the additional Higgs mass, and thus for 
heavy region, that coupling is enhanced. These important differences lead to larger 
regions being consistent with relic density for $M_{\chi}>m_{H_{\rm heavy}}$, when compared to the low mass case where there was only a small patch of 
$M_{\chi}\simeq m_{H_{\rm light}}$ for low mass case. In Fig. \ref{fig:omega-high}, we have again indicated  
$y>1.0$ region in yellow, which lies beyond the perturbative limit. This discards a significant part of 
the relic density allowed parameter space of the model with $M_{\chi}>m_{H_{\rm heavy}}$. We however note that, as the perturbative 
limit on $y$ depends on the cut-off scale of the theory, we can allow a bit more parameter space by being less strict and imposing instead $y>1.16$, by taking smaller values of $\Lambda \sim 10^6$ GeV, which 
is permissible with neutrino mass limits (as detailed in Section \ref{sec:conec}).

\subsubsection{Direct Search}

\begin{figure}[h!]
$$
\includegraphics[height=5.2cm]{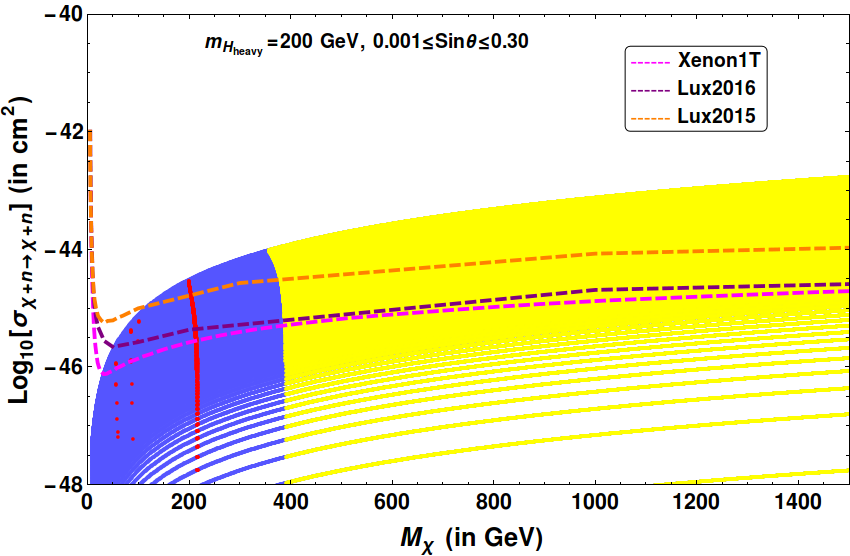}
\includegraphics[height=5.2cm]{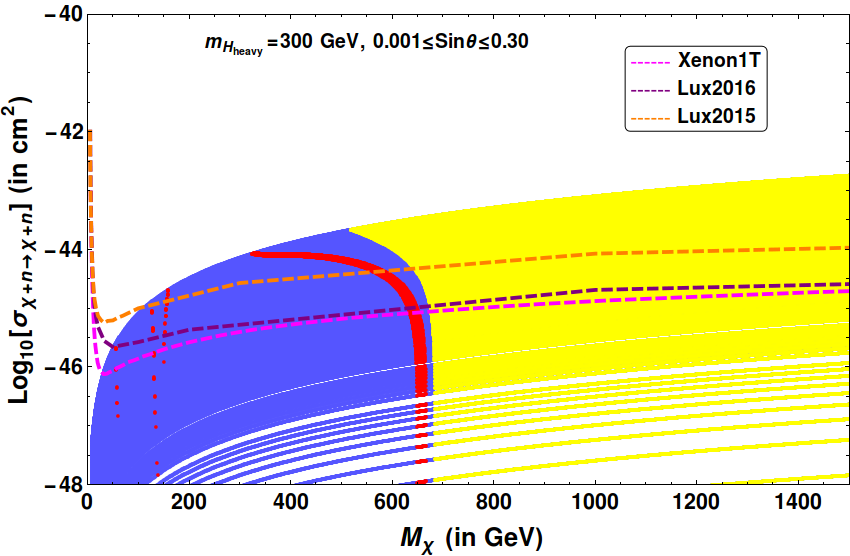}
$$
$$
\includegraphics[height=5.2cm]{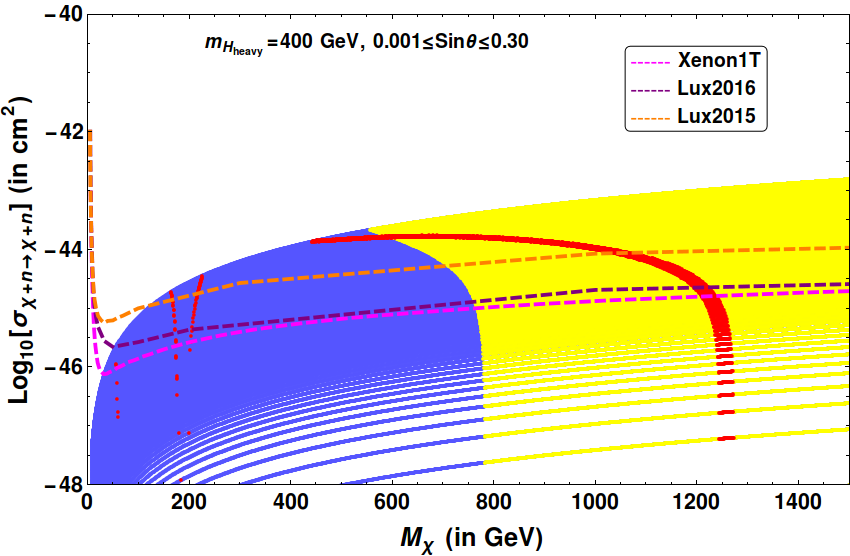}
\includegraphics[height=5.2cm]{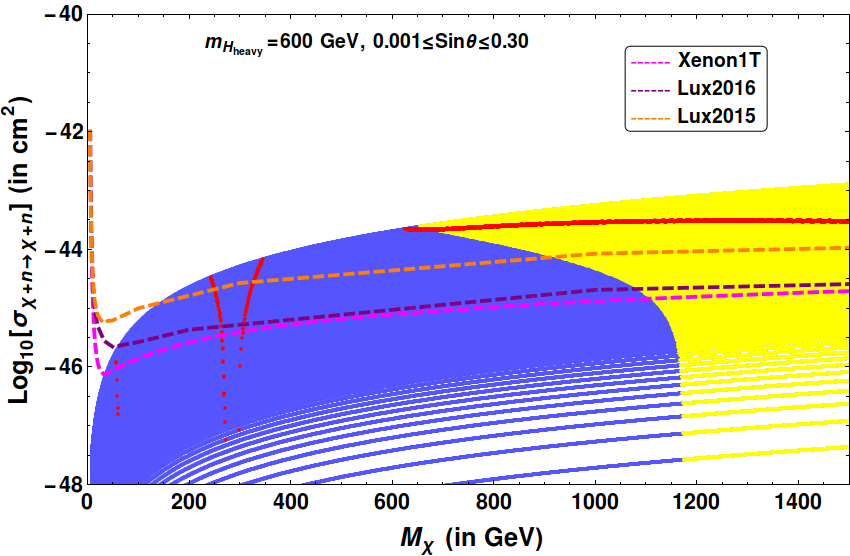}
$$
\caption{Spin independent direct  
search cross section as a function of DM mass ($M_{\chi}$) for 
$m_{H_{\rm Heavy}}$=200, 300, 400 and 600 GeV. The mixing angle is varied between 
$0.001\leq \sin\theta \leq 0.3$. Red dots in respective panels additionally 
satisfy relic density constraints. Bounds from XENON1T and LUX data are shown.  
The yellow region corresponds to $y>1.0$ which may violate the perturbative limit.}
\label{fig:direct-high}
\end{figure}

The variation of the spin independent direct search scattering cross-section with the DM mass $M_{\chi}$
is plotted in Fig. \ref{fig:direct-high} for heavy Higgs masses: $m_{H_{\rm 
heavy}}$ = 200, 300, 400 and 600 GeV. The blue region is obtained by scanning all the 
values of mixing angle in the range of $0.001\leq\sin\theta \leq 0.3$. The red 
points additionally satisfy relic density constraints for the choice of the specific heavy 
Higgs mass. Experimental direct search constraints from LUX and XENON data are shown by the 
dotted lines. Contrary to the low mass region, we see that a significant region of parameter space can 
be found below the direct search region for a wide range of DM masses. When considering 
the points allowed by the relic density, we see that for $M_\chi \gg m_{H_{\rm heavy}}$ we can satisfy 
relic density and direct search bounds in this model outside of the resonance regions. This is due to the freedom 
in choosing a large range of $\sin\theta$ and the annihilation to Higgs final states, which is not constrained 
by the direct search cross-sections. Obviously, for larger $m_{H_{\rm heavy}}$, the required DM 
mass to satisfy relic density and direct search bounds is also larger.  Unfortunately, 
for $m_{H_{\rm heavy}} \gtrsim$ 300 GeV, these points mostly fall into the 
yellow shaded region corresponding to $y>1.0$, which is discarded by perturbative limit.

\begin{figure}[h]
$$
\includegraphics[height=3.8cm]{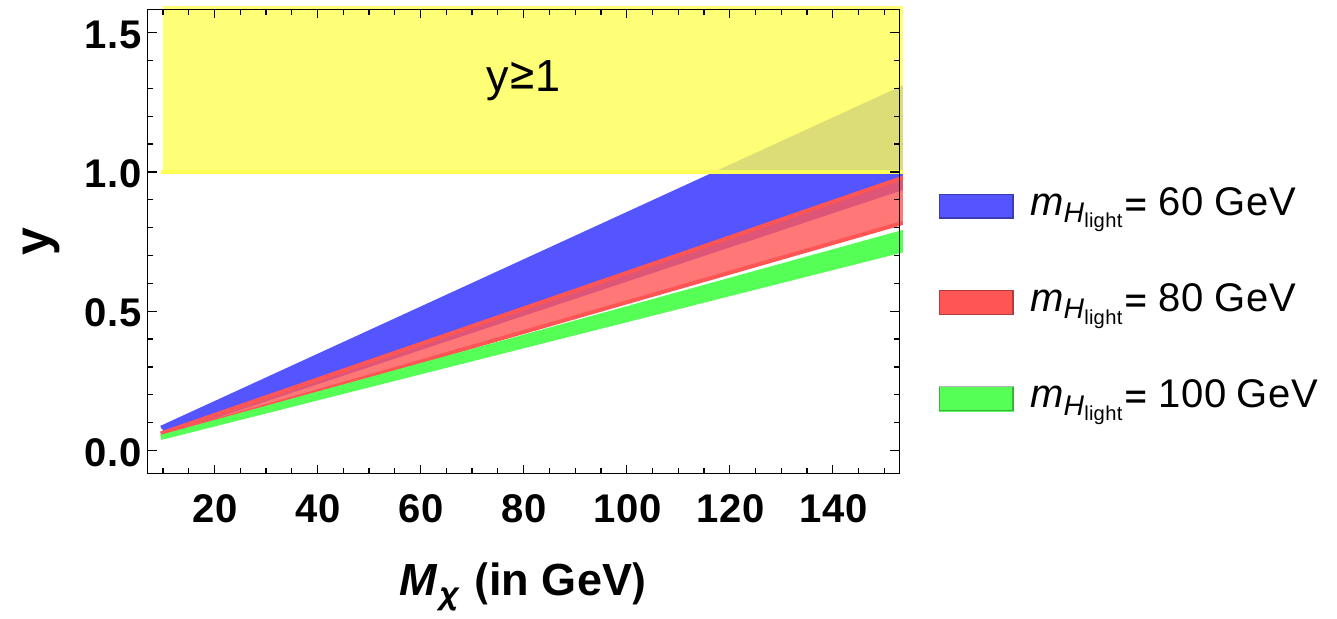}
\includegraphics[height=3.8cm]{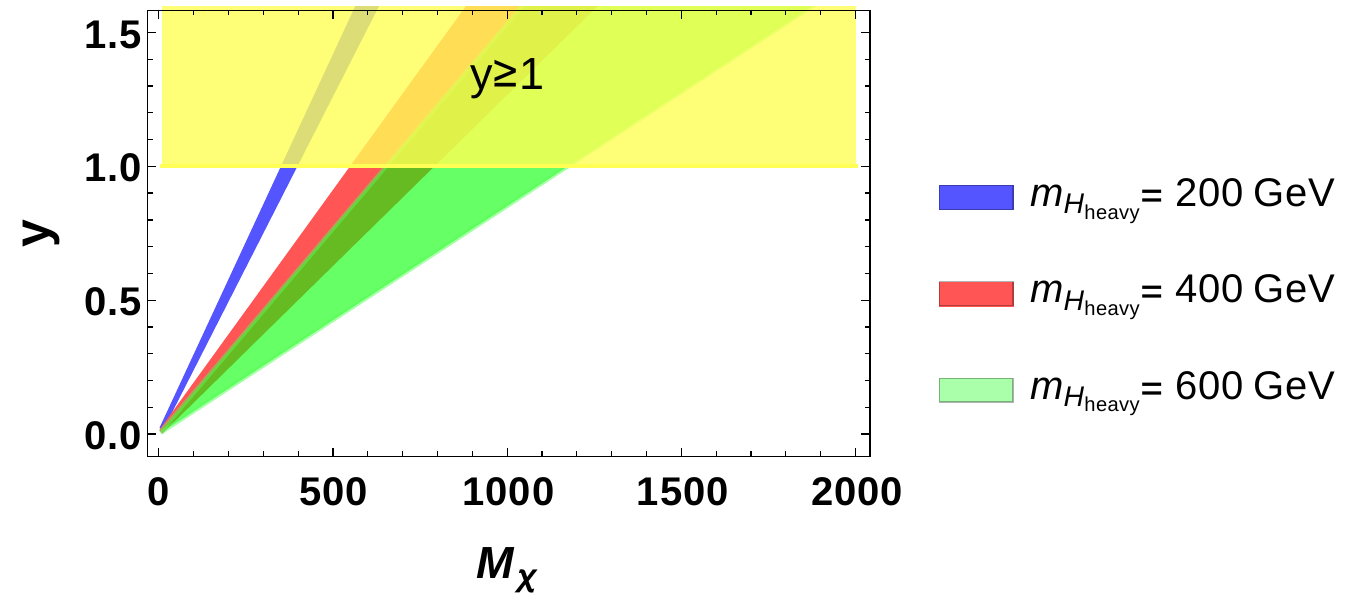}
$$
\caption{Left: Coupling ($y$) vs DM mass $M_{\chi}$ in the low mass region for 
$m_{H_{\rm light}}$=60, 80 and 100 GeV with $0.90\leq \sin\theta \leq 0.999$.
 Right: $y$ vs $M_{\chi}$ for high mass region with $m_{H_{\rm heavy}}$=200, 400, 600 GeV 
for $0.001\leq \sin\theta \leq 0.3$.  The yellow region corresponds to $y>1.0$ which may violate the perturbative limit.}
\label{fig:y}
\end{figure}

One can easily estimate the required coupling $y$ 
given a specific choice of the heavy/light Higgs mass. 
In Fig. \ref{fig:y} we plot $y$ vs $M_{\chi}$ using the relation $y=M_{\chi}/u$ for 
both low mass (Eq. \ref{l1low}) and high mass (Eq. \ref{l1high}) regions 
with the assumption $\lambda=\lambda_1=\lambda_2$. We show the cases of 
the same choices of the benchmark values of the heavy and light Higgs masses 
for both high and low mass regions as used in the DM analysis, spanning the admissible range of mixing angle. 
We see that for larger DM mass, the required coupling $y$ is larger. 
Also for smaller BSM Higgs mass, the required coupling $y$ is larger given a specific DM mass.
The perturbative disallowed region with $y>1.0$ is shown by the yellow band.
Therefore, larger values of DM masses  are disfavoured by the high coupling $y$ 
specifically when the BSM Higgs mass is chosen smaller. The implication
is that in this model, viable DM masses are mainly those in the resonance regions (as has already been discussed above). 
For the low mass region, points with $M_\chi \simeq m_{H_{\rm light}}$ allowed by the DM constraints
have admissible value for $y$ and therefore can be considered, but for the 
high mass region, allowed points with $M_\chi \gg m_{H_{\rm heavy}}$ are 
disfavoured by imposing $y<1$. Furthermore, for $M_\chi < m_{h_{\rm SM}}/2$ Higgs invisible decays constraints severely restrict the allowed values of $\sin \theta$ (see Appendix \ref{apnd3}).

\begin{table}[h]
\centering
\centering
\resizebox{15cm}{!}{%
\begin{tabular}{|c|c|c|c|c|c|c|c|c|}\hline
  Mass of the Scalar  & $\sin\theta$ & $\lambda$ & $\lambda_{12}$ & $u$ GeV & 
$y$ & $M_{\chi}$ GeV &  $\Omega h^2$ & Log$_{10}$[$\sigma_{\rm SI}$]$\times 
10^{-45}$ cm$^2$ \\\hline
\multirow{4}{*}{$m_{H_{\rm light}}=$100 GeV} & 0.993 & 0.13 & 0.019  & 196.99 & 
0.236  & 46.59 & 0.118 & -46.10 \\
                  & 0.996 & 0.13 & 0.011 & 196.45 & 0.292 & 57.46 & 
0.121 & -46.15 \\
                   & 0.998 & 0.13 & 0.008 & 196.08 & 0.496 & 97.34 & 0.121 & 
-45.99   \\
                  & 0.999 &  0.13 & 0.005 & 195.89 & 0.497 & 97.40 & 0.119 & 
-46.29   \\ \hline
\multirow{3}{*}{$m_{H_{\rm heavy}}=$200 GeV} 
          & 0.069  & 0.133  & 0.017 & 389.43 & 0.154 & 60 & 0.122  &  -46.90\\ 
          & 0.031 & 0.131  & .008 & 391.01 & 0.233 & 91 & 0.119  & -47.23 \\ 
          & 0.011 & 0.131 & 0.003  & 391.35 & 0.552 & 216 & 0.121  &  -47.37\\ 
\hline
\multirow{2}{*}{$m_{H_{\rm heavy}}=$400 GeV} 
         & 0.129 & 0.150 & 0.104 & 723.90 & 0.083 &  60 & 0.120 & -46.55 \\
         & 0.115 & 0.146 & 0.091 & 735.04 & 0.278 & 204  & 0.118 & -45.60 
\\\hline
\multirow{2}{*}{$m_{H_{\rm heavy}}=$600 GeV} 
         & 0.165 & 0.208 & 0.248  & 918.13 & 0.065 & 60 & 0.119 & -46.50 \\
         & 0.093 & 0.155  & 0.121 & 1072.67  & 0.283 & 304 & 0.118  & -45.70 \\
\hline
\end{tabular}
}\
\caption{\label{tab:bmps}{Some characteristic benchmark points for the model in both low and high mass regions 
satisfying dark matter relic density and direct search constraints.}}
\end{table}
In table \ref{tab:bmps}, we have listed some benchmark points from both low and 
high mass regions that are allowed by DM phenomenology and fulfil the perturbative limit on $y$. The 
benchmark points lie essentially in the two Higgs resonance regions as discussed. Also it is permissible to have 
DM mass in the vicinity of the light/heavy Higgs mass (when particularly the heavy Higgs is not too heavy, in order to avoid the perturbative limit).

\section{Seesaw mechanism and the connection to dark matter}\label{sec:conec}

 \begin{figure}[h!]
 $$
 \includegraphics[height=5.5cm]{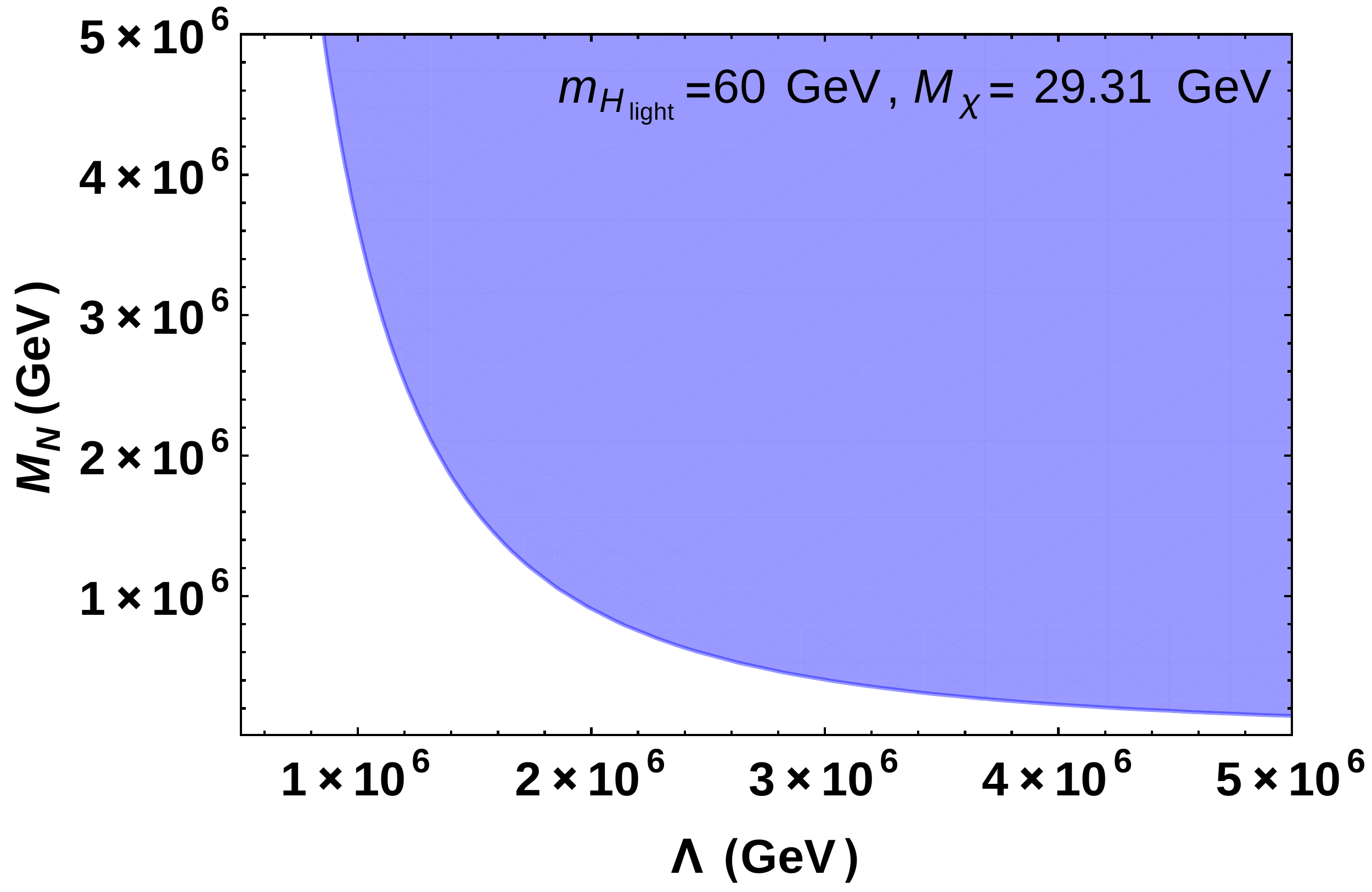}
 \includegraphics[height=5.5cm]{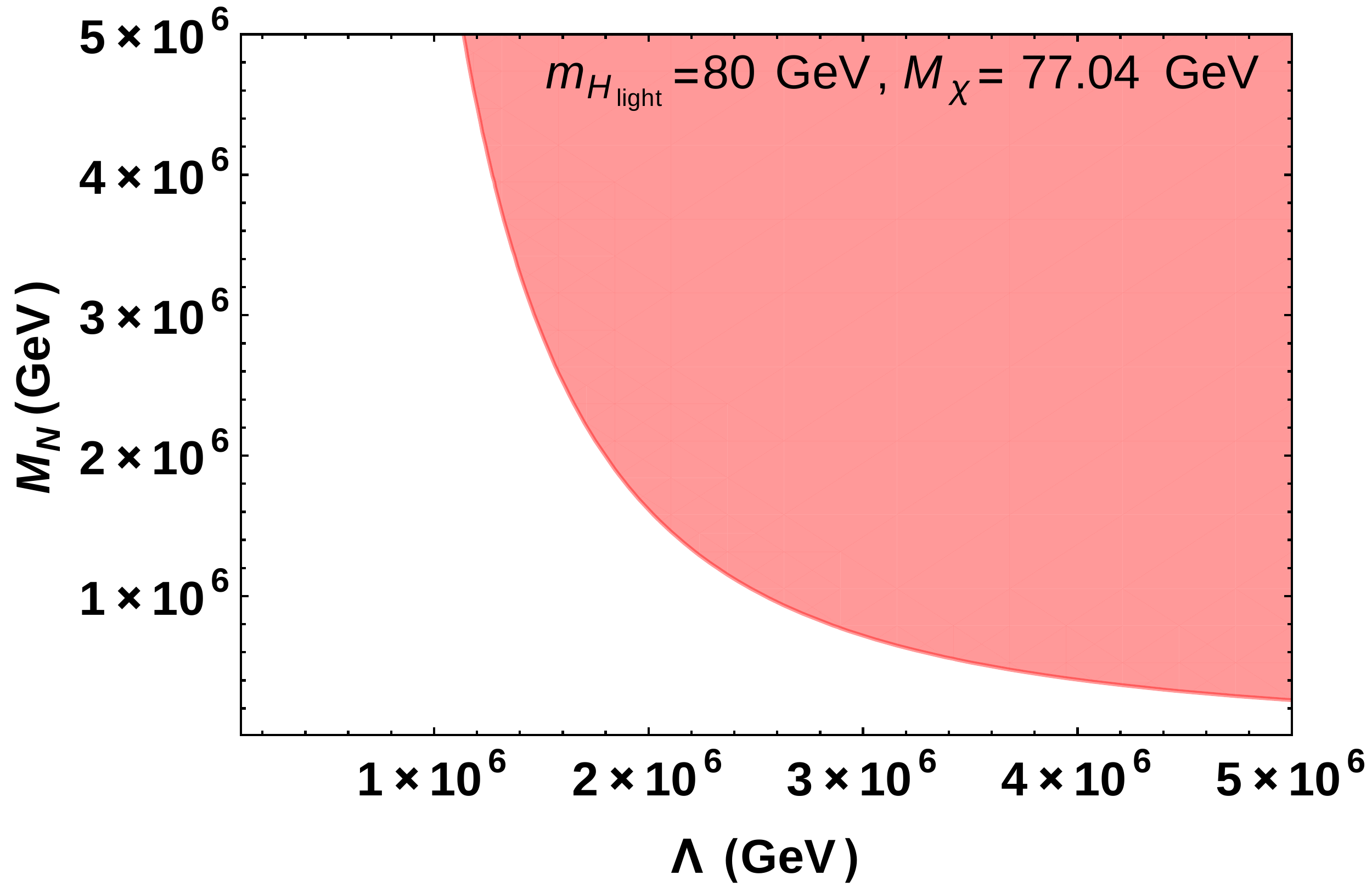}
 $$
 $$
 \includegraphics[height=5.5cm]{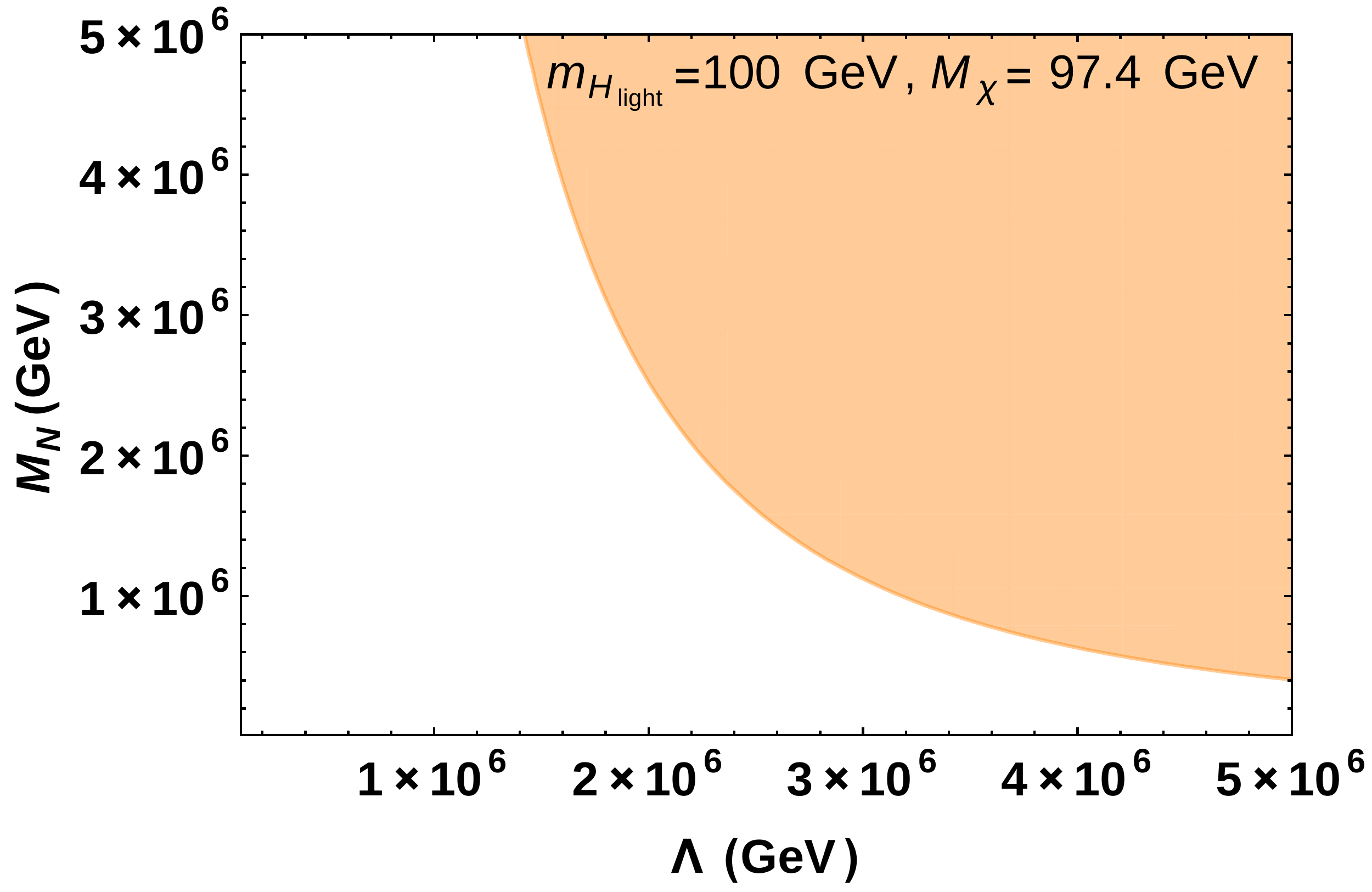}
 $$
 \caption{Correlation between right handed neutrino mass $M_N$ and cut-off 
scale 
 $\Lambda$ for $m_{H_{\rm light}}$ = 60, 80 and 100 GeV with corresponding DM 
masses $M_\chi$=29.31, 77.04 and 97.4 GeV respectively  satisfying  both relic 
density and direct search constraints.  \label{fig:mnu1}
}
 \end{figure}

As has already been mentioned, beyond generating the dark matter mass, the 
scalar field $\phi$ is also instrumental for generating the Yukawa couplings of the neutrinos which then lead to the
light neutrino mass 
through type-I seesaw. From Eq. (\ref{mnu}), the seesaw formula in the present 
scenario can be written as 
\begin{eqnarray}\label{mnu2}
 m_{\nu}&=&\frac{v^2 u^2}{\Lambda^2 
M_N}=\frac{v^2}{y^2}\frac{(M_{\chi}/\Lambda)^2}{M_N}.
\end{eqnarray}
Recent cosmological observation by Planck suggests that sum of absolute masses of three 
light neutrinos to be $\sum m_i\leq 0.23$ eV~\cite{Ade:2015xua}. 
Using Eq. (\ref{mnu2}), the bound on the masses of the light neutrinos can be written as 
\begin{equation}\label{mnubound}
 \frac{v^2}{y^2}\frac{M_{\chi}^2}{\Lambda^2 M_N}\leq 0.23 ~{\rm{eV}}. 
\end{equation}

Therefore, once we have an idea of the DM mass from 
the relic density and direct search constraints, as we have already obtained in 
the previous section, we employ Fig. \ref{fig:y} (left and right panels for low and high mass 
regions respectively) to determine the corresponding Yukawa $y$. 
The allowed regions for the right handed neutrino mass ($M_N$) and cut-off scale 
of the theory ($\Lambda$) can then 
be obtained from the above constraint on the combination $\Lambda^2 M_N$ 
following Eq. (\ref{mnubound}). For example, in the low mass region, from Fig. \ref{fig:direct-low} we find that with 
$m_{H_{\rm light}}=60,80$ and 100 GeV both relic density and direct search 
constraint can be satisfied for DM mass $M_\chi$= 29.31, 77.04 and 97.4 GeV 
respectively. Hence following Eq. \ref{mnubound}, we draw a correlation between 
the right handed neutrino mass $M_N$ and cut-off scale $\Lambda$ in Fig. 
\ref{fig:mnu1}. The shaded regions in all the panels are the allowed regions by 
all constraints. This indicates to the choice of a cut off scale of the model 
$\Lambda \gtrsim 10^6$ GeV as has already been advocated. 
This in turn then implies that the perturbative limit on the coupling $y$ to be as stringent as 
$y<1.16$,  which we have imposed in the DM analysis (see Appendix A for details). We also note that 
in the top left figure, DM mass is less than half of SM Higgs mass, which causes the SM Higgs to decay 
invisibly. The LHC data constrains such a case to limit the mixing $\sin\theta$ (see Appendix C for details), 
which falls within our chosen range.

\begin{figure}[h]
 $$
 \includegraphics[height=5.5cm]{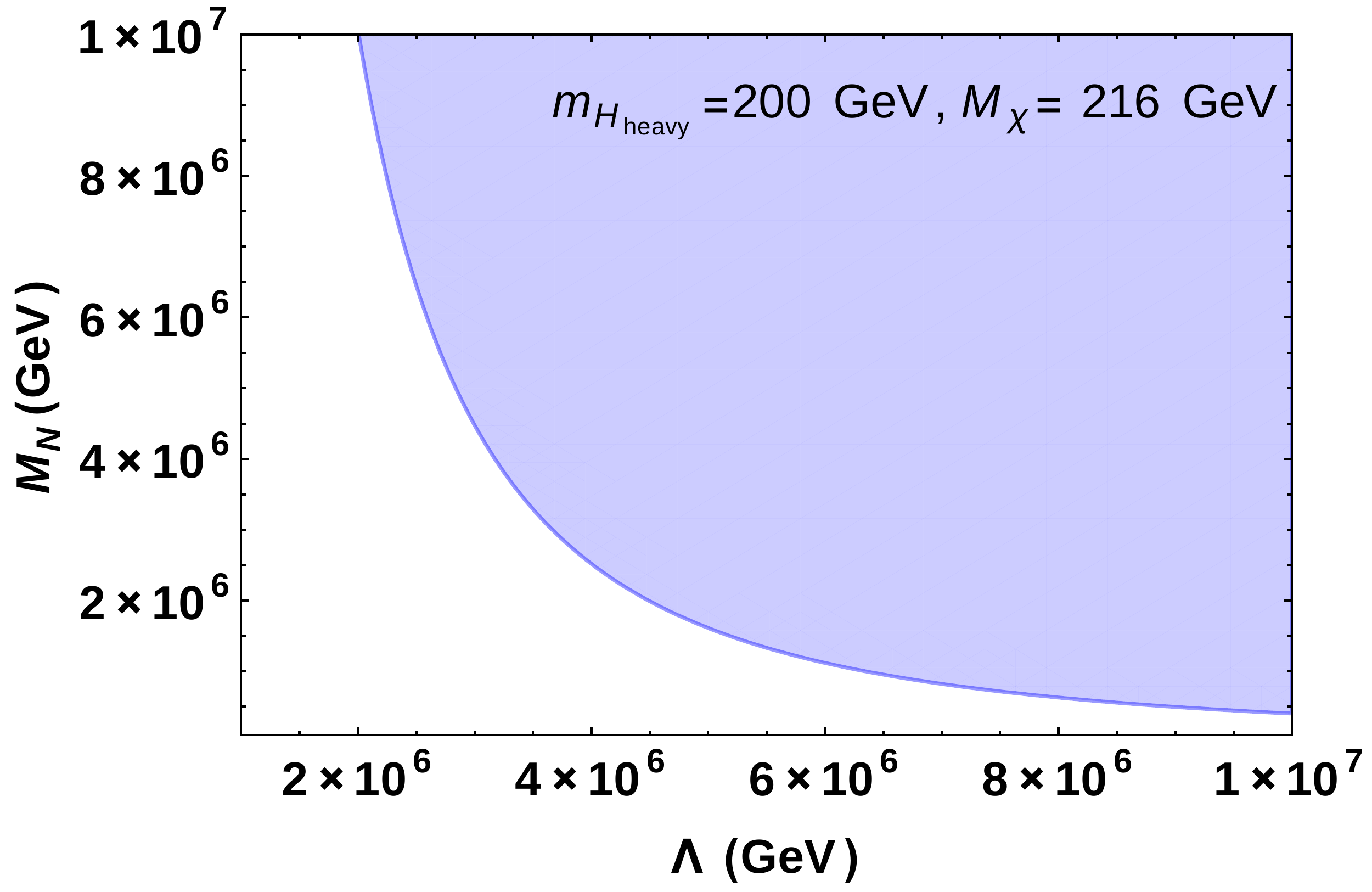}
 \includegraphics[height=5.5cm]{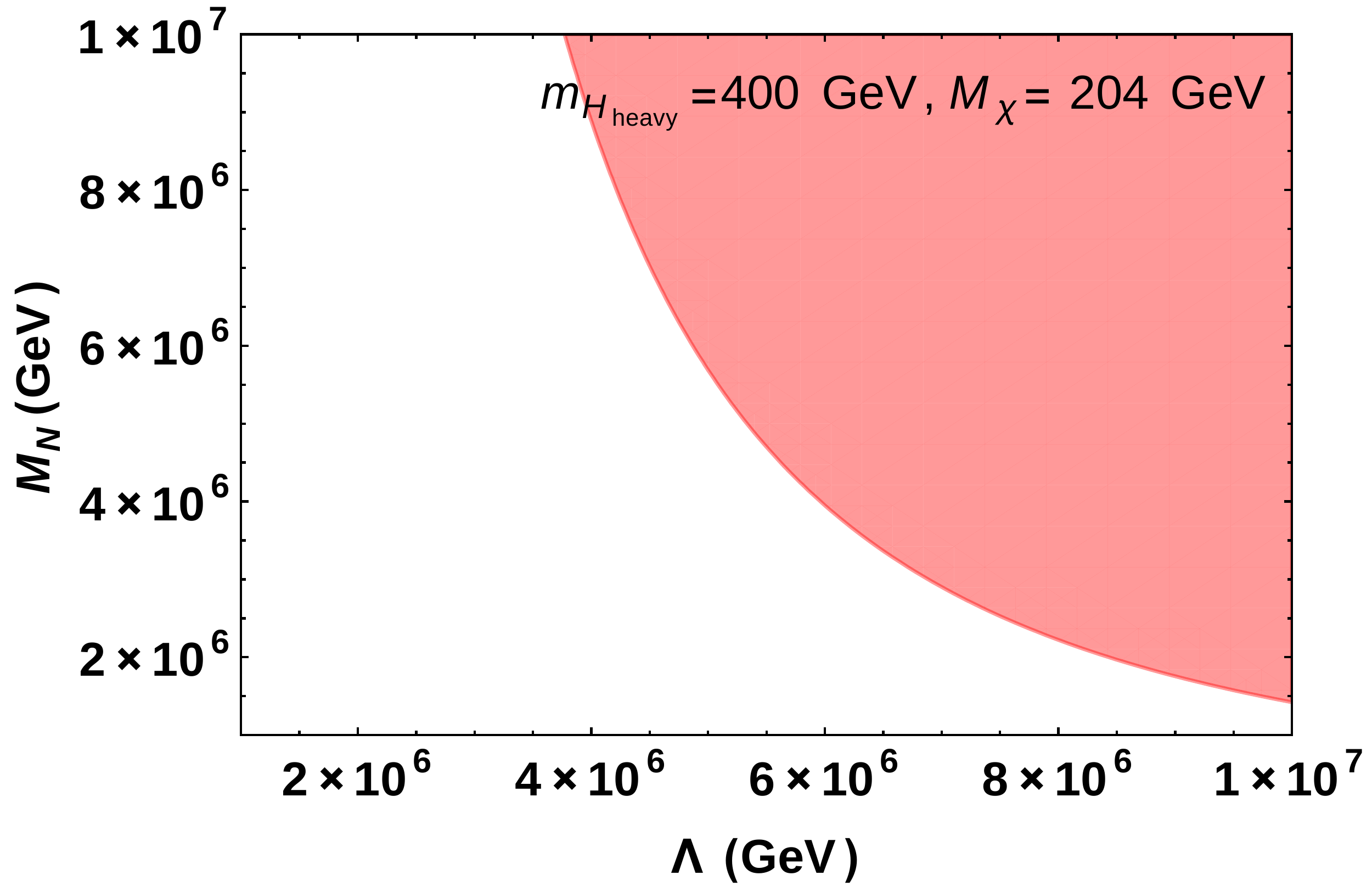}
 $$
  $$
 \includegraphics[height=5.5cm]{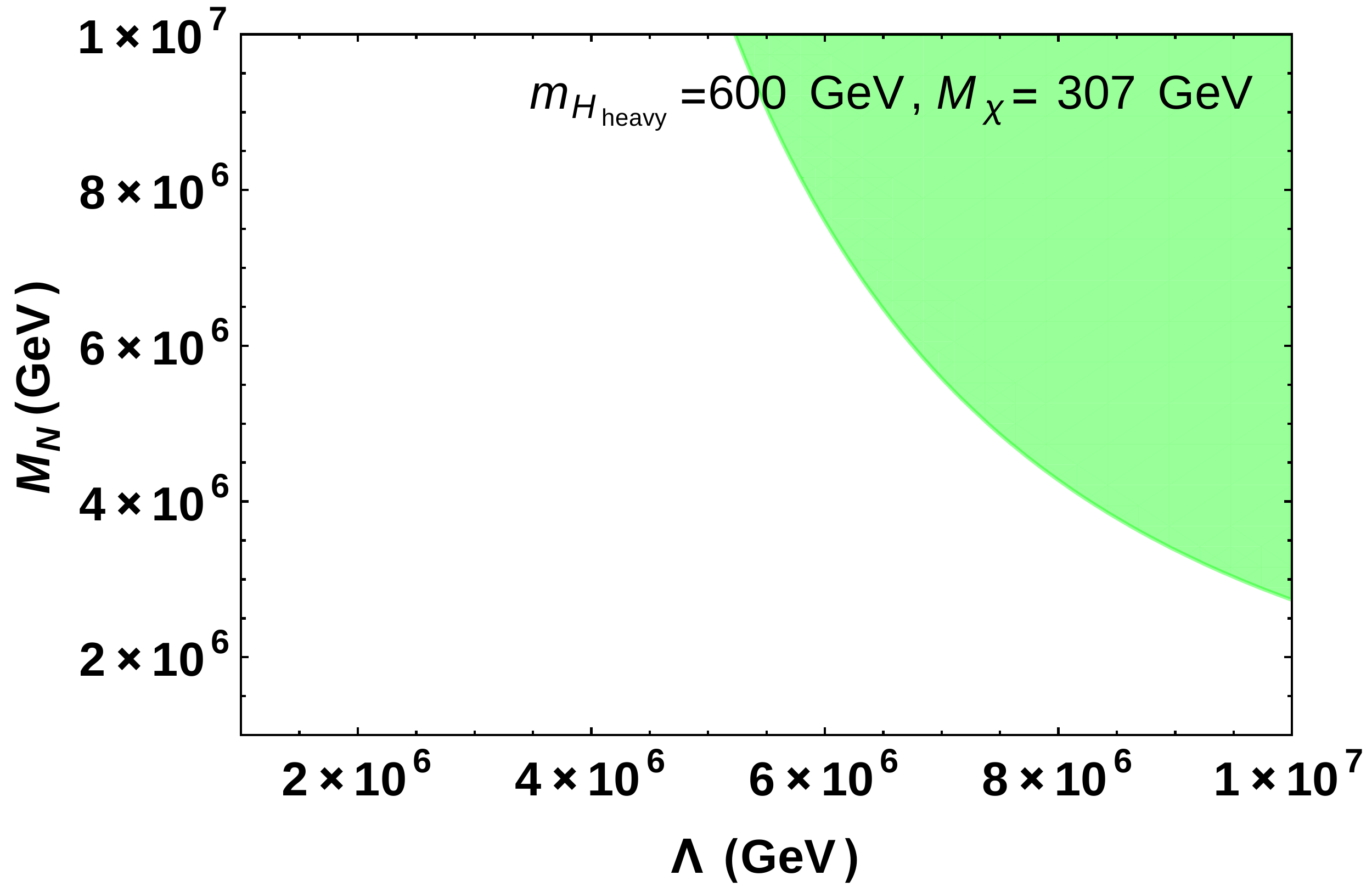}
 $$
 \caption{Correlation between right handed neutrino mass $M_N$ and cut-off 
scale $\Lambda$ for successful generation of light neutrino masses for different choices of heavy Higgs mass 
 $m_{H_{\rm heavy}}$ = 200, 400, 600 GeV, which points out to certain definite choices of 
 DM masses as $M_{\chi}$ = 216, 204, 307 GeV respectively (corresponding to 
$\sin\theta=0.011, ~0.115, ~0.119$) to satisfy relic density and direct search constraints together with perturbative bounds.}
 \label{fig:mnu2}
 \end{figure}

Similarly, for high mass region, we draw the similar correlations between right 
handed neutrino mass and the cut-off scale for different choices of Heavy Higgs 
mass which unambiguously point out to specific DM masses to satisfy relic 
density and direct search constraints as given in Fig. \ref{fig:mnu2}. Here 
we have drawn the correlations for DM masses 
$M_{\chi}$ = 216, 204, 307 GeV respectively (corresponding to 
$\sin\theta=0.011, ~0.115, ~0.119$). For heavier DM mass, the right handed neutrino mass and cut-off scale are also required to be heavier.

\begin{figure}[h]
 $$
 \includegraphics[height=6.1cm]{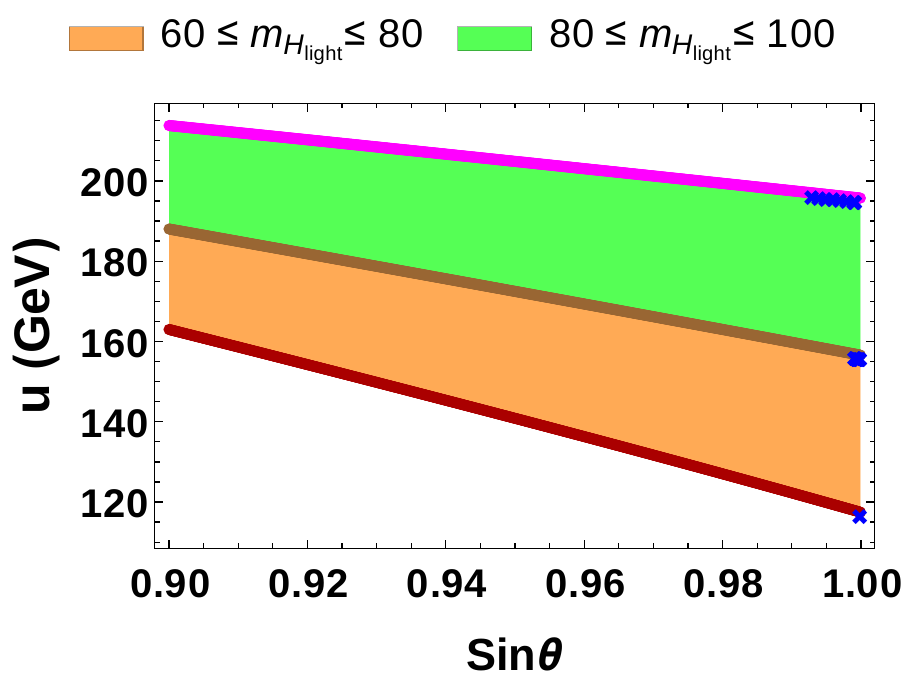}
 \includegraphics[height=5.2cm]{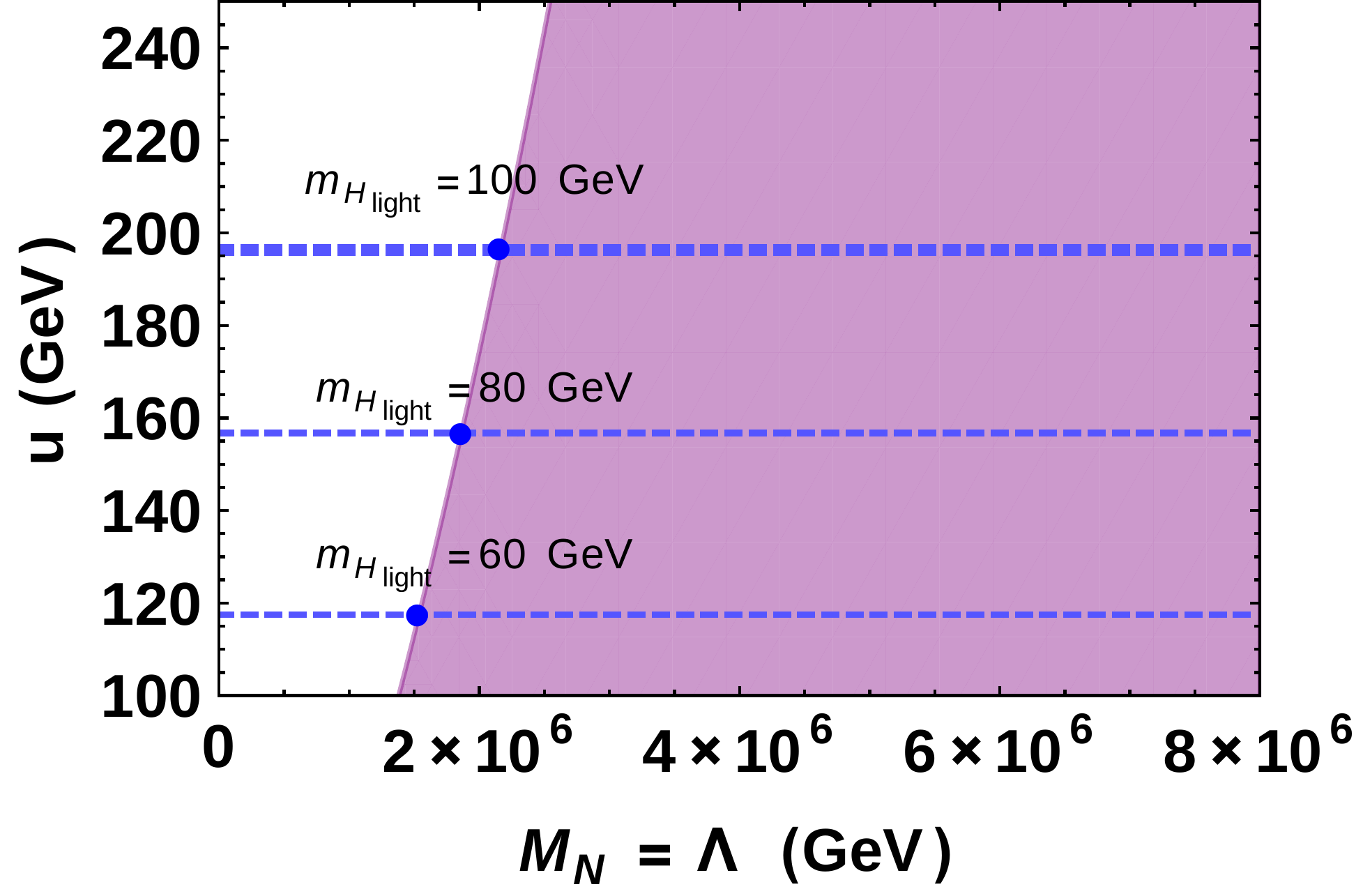}
 $$
 \caption{Left panel: $u (=M_{\chi}/y)$ vs $\sin\theta$ plot for 
$\lambda_1=\lambda_2=\lambda$ for various range of $m_{H_{\rm light}}$. Here 
the magenta, brown and dark red dots represent allowed points obtained (from 
Fig. \ref{fig:omega-1ow}) relic density for $m_{H_{\rm light}}=100$, 80 
and 60 GeV. Blue dots on each line are the points allowed by the direct search limit. Right panel:  $u (=M_{\chi}/y)$ as a function of 
$\Lambda$ (or $M_{N}$) considering $M_N = \Lambda$ in Eq. \ref{mnu2} and 
\ref{mnubound}. Purple shaded region represents $m_{\nu}\leq 0.23$ eV with $M_N 
=\Lambda$ and the horizontal blue patches represent the allowed region of $u 
(=M_{\chi}/y)$ obtained the left panel for the specific light Higgs masses. This imposes a stringent constraint on 
the lower limit of the cut-off scale $\Lambda$ (and RH neutrino mass $M_{N}$).}
 \label{fig:c1}
 \end{figure}

We consider now in more detail the correlation between neutrino 
sector and dark matter sector. In the left panel of Fig. \ref{fig:c1} we have 
plotted  $u (=M_{\chi}/y)$ vs $\sin\theta$ for 
$\lambda_1=\lambda_2=\lambda$ for ranges of $m_{H_{\rm light}}$. Here 
the magenta, brown and dark red dots represent allowed points satisfying only 
dark matter relic density (obtained from Fig. \ref{fig:omega-1ow})
for $m_{H_{\rm light}}=100$, 80 and 60 GeV respectively. The blue dots overlaid on 
each of those 3 lines further satisfy the direct search limit obtained from Fig. 
\ref{fig:direct-low}. Here we find that only $\sin\theta$ values rather close to 1 
satisfy the DM constraints. This plot gives an 
estimate for the scalar singlet vev, $u$, for each low mass case. In 
the right panel of Fig. \ref{fig:c1},  we have plotted $u (=M_{\chi}/y)$ as a 
function of $\Lambda$ (or $M_{N}$) considering $M_N = \Lambda$ in Eq. \ref{mnu2} 
and \ref{mnubound}. The purple shaded region represents allowed parameter space 
in the $u(=M_{\chi}/y)$-$\Lambda(=M_{N})$ plane satisfying upper bound on the 
sum of light neutrino mass $m_{\nu}\leq 0.23$ eV with $M_N = \Lambda$. Here the 
horizontal blue patches represents the allowed region of $u (=M_{\chi}/y)$ 
obtained in the left panel (following the analysis of dark matter sector in the 
previous section). This imposes a stringent constraint on the lower limit of 
the cut-off scale $\Lambda$ (and RH neutrino mass). Hence for the low mass 
region (with $m_{H_{\rm light}}=100$, 80 and 60 GeV), we find the lower limit 
on $\Lambda (M_N)$ to be $\Lambda (M_N) \geq 2.2\times10^6$, $1.8\times10^6$ 
and $1.5\times10^6$ GeV respectively.

\begin{figure}[h]
 $$
 \includegraphics[height=6.1cm]{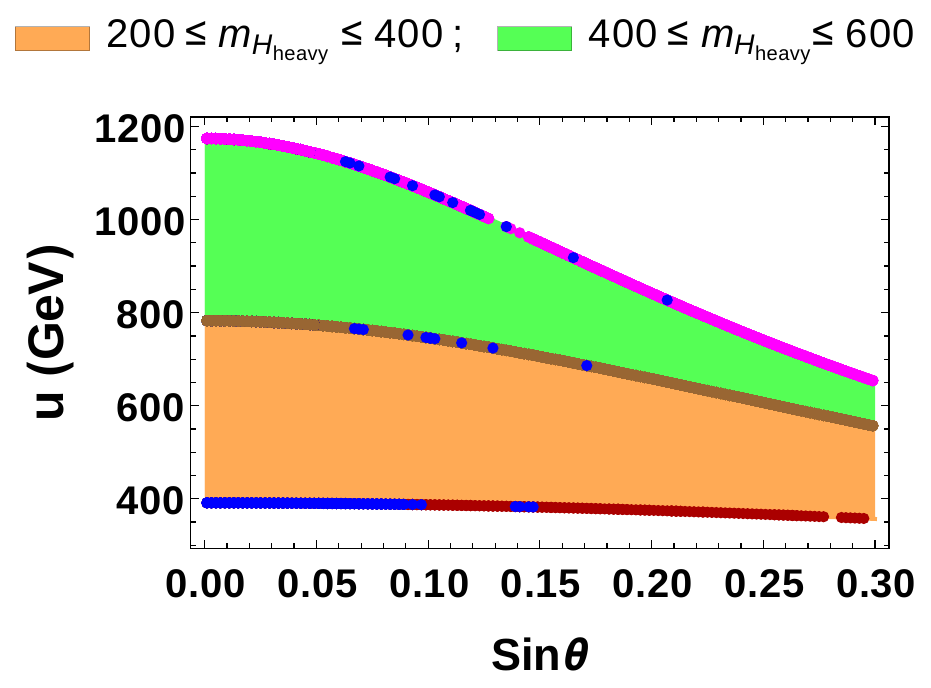}
 \includegraphics[height=5.2cm]{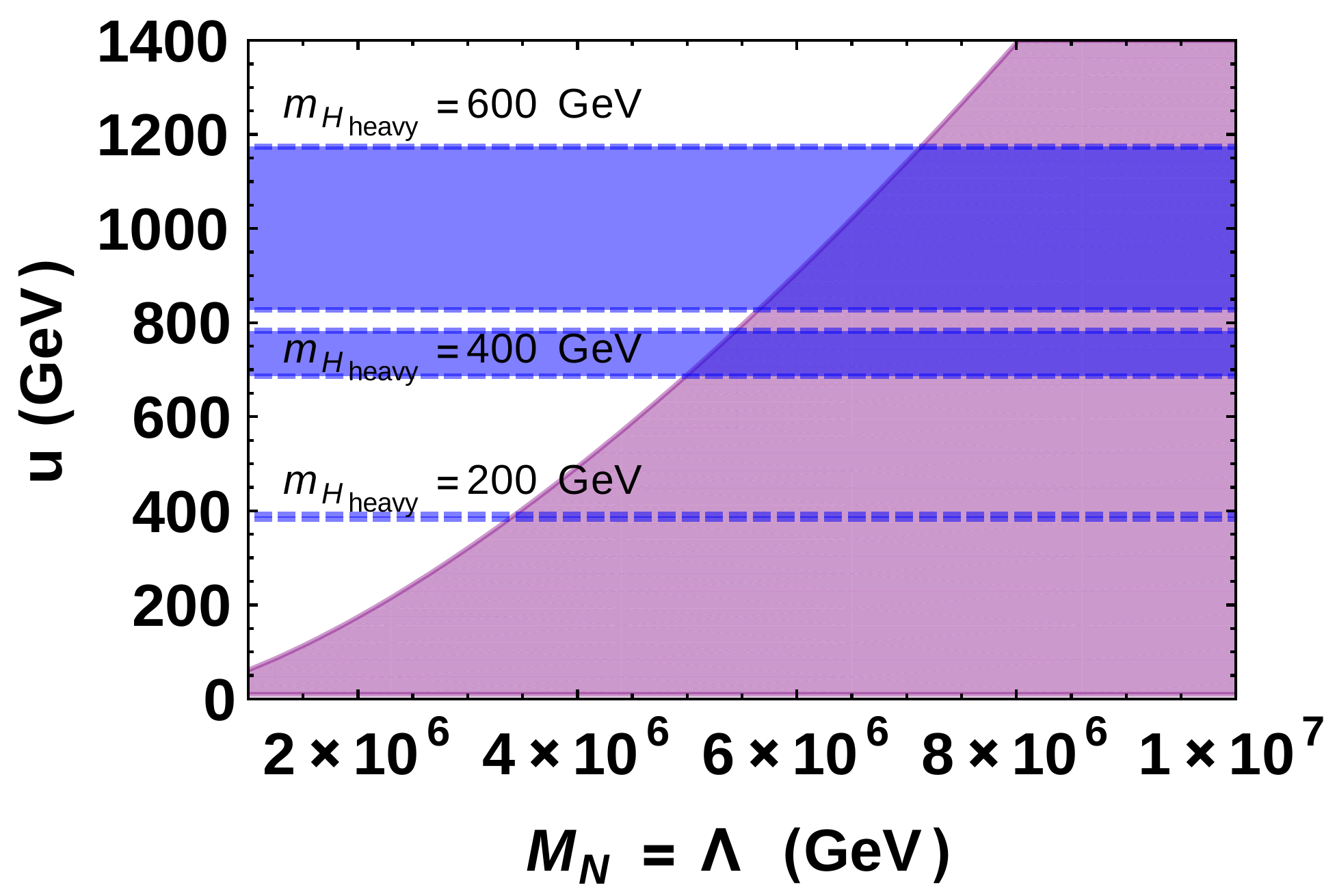}
 $$
 \caption{Left panel: $u (=M_{\chi}/y)$ vs $\sin\theta$ plot for 
$\lambda_1=\lambda_2=\lambda$ for  $ 200 \leq m_{H_{\rm heavy}} \leq 400$ GeV 
(orange shaded region) and $ 400 \leq m_{H_{\rm heavy}} \leq 600$ GeV 
(green shaded region) respectively. Here the magenta, brown and deep red dots 
represent allowed points obtained (from 
Fig. \ref{fig:omega-high}) relic density only, for $m_{H_{\rm light}}=100$, 80 
and 60 GeV. Blue dots on each line are actual allowed points which also 
satisfy direct search limit. Right panel:  $u (=M_{\chi}/y)$ as a function of 
$\Lambda$ (or $M_{N}$) considering $M_N = \Lambda$ in Eq. \ref{mnu2} and 
\ref{mnubound}. Purple shaded region represents $m_{\nu}\leq 0.23$ eV with $M_N 
= \Lambda$ and the horizontal blue patch represents the allowed region of $u 
(=M_{\chi}/y)$ obtained the left panel for the specific heavy Higgs masses. This imposes a constraint on 
the lower limit of the cut-off scale $\Lambda$ (and RH neutrino mass $M_{N}$).}
 \label{fig:c2}
 \end{figure}

A similar analysis can be performed for the high mass 
region. In the left panel of Fig. \ref{fig:c2}, we have again 
plotted $u (=M_{\chi}/y)$ against $\sin\theta$ for 
$\lambda_1=\lambda_2=\lambda$ 
for various range of $200 \leq m_{H_{\rm heavy}} \leq 400$ GeV and $400 \leq 
m_{H_{\rm heavy}} \leq 600$ GeV as depicted by orange and green shaded regions. 
In this panel the magenta, brown and dark red dots represents the allowed 
points satisfying correct dark matter relic density as given in Fig. 
\ref{fig:omega-high} for $m_{H_{\rm heavy}}=200$, 400 and 600 GeV 
respectively. Blue dots additionally satisfy direct search 
constraints as obtained following Fig. \ref{fig:direct-high}. Here 
a relatively large 
region of $\sin\theta$ satisfies all the DM constraints representing a wide 
range for scalar singlet vev $u$. In the right panel of Fig. \ref{fig:c2}, we 
have again plotted $u (=M_{\chi}/y)$ as a function of $\Lambda$ (or $M_{N}$) 
considering $M_N = \Lambda$ in Eq. \ref{mnu2} and \ref{mnubound} for the high 
mass region. The purple shaded region represents allowed parameter 
space in the $u(=M_{\chi}/y)$-$\Lambda(=M_{N})$ plane satisfying the upper bound on 
the sum of light neutrino mass $m_{\nu}\leq 0.23$ eV with the 
consideration $M_N = \Lambda$. In this right panel, horizontal shaded regions 
illustrated in blue (for $m_{H_{\rm heavy}}=200$, 400 and 600 
GeV) represent the allowed regions for $u (=M_{\chi}/y)$ obtained 
from the left panel satisfying correct dark phenomenology. From the 
intercepting regions once again we can obtain the corresponding lower limit on 
$\Lambda$ (and $M_N$). Here we find a relatively wider lower limit 
for $\Lambda (M_N) \geq 3.4\times10^6$, (5.0-5.4)$\times 10^6$ and 
(5.64-7.1)$\times 10^6$ GeV for $m_{H_{\rm heavy}}=200$, 400 and 600 GeV 
respectively due to due larger allowed region for $\sin\theta$ (and hence 
corresponding $u$ in the left panel). 

 Now if we compare the left panels of Fig. \ref{fig:c1} 
and \ref{fig:c2} we find that for low mass region, only $\sin\theta$ values 
close to 1 (denoted by the blue dots for $m_{H_{\rm light}}$= 100, 80 and 60 
GeV respectively) satisfy direct search constraint. Hence the lighter Higgs 
is dominantly scalar singlet and s-channel contribution to the DM annihilation 
almost vanishes and t-channel diagrams dominantly contributes both in relic 
density and direct search constraints. In contrast, for the high mass region the
bound on $\sin\theta$ is a bit relaxed (also denoted by the blue dots for 
$m_{H_{\rm heavy}}$= 200, 400 and 600 GeV respectively) in order to satisfy 
both relic density and direct search constraints. This eventually leads to the 
fact that for low mass region, for a specific value of $m_{H_{\rm light}}$ the 
lower limit for the cut-off scale $\Lambda$ (or RH neutrino mass $M_{N}$) is 
tightly constrained as evident from the right panel of Fig. \ref{fig:c1}. But for the
high mass region, due to the large allowed range for $\sin\theta$, for a fixed value 
of $m_{H_{\rm heavy}}$ we have a wide allowed range for the lower 
limit for the cut-off scale $\Lambda$ (or RH neutrino mass $M_{N}$). This is 
shown in the right panel of Fig. \ref{fig:c2}.

\section{Summary and Conclusions}\label{sec:conc}
We have successfully constructed a framework where seesaw and DM
sectors are related to each other by a common scalar mediator. The relation is mainly 
restricted by the vev of the scalar singlet field which yields both DM mass and 
controls the neutrino Yukawa coupling in the type-I seesaw within an effective 
theory framework. The  first part of the framework described above constrains 
the DM mass from relic density and direct search constraints, depending on the
mass of mixing associated with the additional Higgs. Given this information, we have 
explored the correlation between the right handed neutrino masses with the 
cut-off scale present in the theory. The main success of the set-up is to 
establish the correlation between DM sector and neutrino sector in a coherent 
manner. The scenario naturally accommodates two Higgses, one of which can be 
identified with the Higgs discovered at the LHC. We study both the cases where 
the additional Higgs field (other than the SM one) is heavier and lighter than 
the SM Higgs. We find that with the second Higgs as the lighter than the 
SM Higgs, the allowed DM phenomenology restricts DM $\sim  m_{H_{\rm light}}$.
On the 
heavy Higgs region, this connection is little relaxed as a larger region of 
allowed parameter space is possible for DM.
In either cases, there is a prediction for $M_N$ and $\Lambda$ to be larger than $10^6$ GeV.
For simplicity, here we 
consider the quartic couplings of the two scalars present in 
the theory as same. The analysis can easily be extended for different 
values of these couplings, which would yield an extended parameter space 
(as hinted in the Appendix \ref{apnd2}).

Very importantly we note that as the coupling $y$ of the scalar mediator to the DM in this model 
is determined directly by DM mass, the perturbative limit on this coupling
reduces the allowed parameter space of the model. The limit on the 
cut off scale $\Lambda \gtrsim 10^6$ GeV,  indicates that $y\le 1.16$ to be within perturbative limit. 
This explicitly excludes very high DM masses, which in the heavy Higgs region would otherwise satisfy both 
relic density and direct search constraints. This therefore enhances the predictivity of the model.

As we have already stated, the choice of the dark sector was
chosen as a specific example  only, and one may do a similar model building 
exercise to connect seesaw 
mechanism to some other 
DM sector. On the other hand, the correlation requires 
the knowledge of the heavy or light Higgs mass and its mixing with the SM Higgs doublet, which is 
difficult to find at the current status of collider search experiment. As our framework involves the 
two heavy scales, namely the RH neutrino mass $M_N$ and the cut-off scale $\Lambda$, it would be interesting to 
find an UV complete construction, although this is beyond the scope of the 
current work  and requires involvement of more fields and symmetry.

\section*{Acknowledgements}

SB is supported by DST- INSPIRE Faculty grant  IFA-13 PH-57 at IIT Guwahati.
IdMV acknowledges funding from Funda\c{c}\~{a}o para a Ci\^{e}ncia e a 
Tecnologia (FCT) through the contract IF/00816/2015, partial support by 
Funda\c{c}\~ao para a Ci\^encia e a Tecnologia (FCT, Portugal) through the 
project CFTP-FCT Unit 777 (UID/FIS/00777/2013) which is partially funded through 
POCTI (FEDER), COMPETE, QREN and EU, and partial support by the National Science 
Center, Poland, through the HARMONIA project under 
contract UMO-2015/18/M/ST2/00518 (2016-2019). B.\,K. acknowledges hospitality at 
University of Southampton where this work was initiated. S.\,F.\,K. 
acknowledges the STFC Consolidated Grant ST/L000296/1 and the  European Union's 
Horizon 2020 Research and Innovation programme under Marie Sk\l{}odowska-Curie 
grant agreements Elusives ITN No.\ 674896 and InvisiblesPlus RISE No.\ 690575. 

\appendix

\section{Perturbative limit on coupling $y$}
\label{apnd1}

With the Lagrangian term for the interaction between the dark matter fermion and the scalar given by:
\begin{equation}
\nonumber
\mathcal{L}_{Yuk} = -y \phi \bar\chi^c \chi~,
\end{equation}
the perturbative limit on $y$ at any scale is expected to satisfy $y<\sqrt{4\pi}$. Hence, we should have 
$y(\Lambda) < \sqrt{4\pi}$, where $y(\Lambda)$ is the coupling at the cut off scale $\Lambda$ 
of the theory. Now it turns out from the 
simplified analysis (where we take $\Lambda = M_N$) that $\Lambda$ has a lower bound, $\Lambda \gtrsim 10^6$ GeV, obtained mainly 
from neutrino mass limit with a suitable heavy/light Higgs mass to satisfy the DM constraints (see Figs. 11 and 12). 
\begin{figure}[h]
$$
\includegraphics[height=7.5cm]{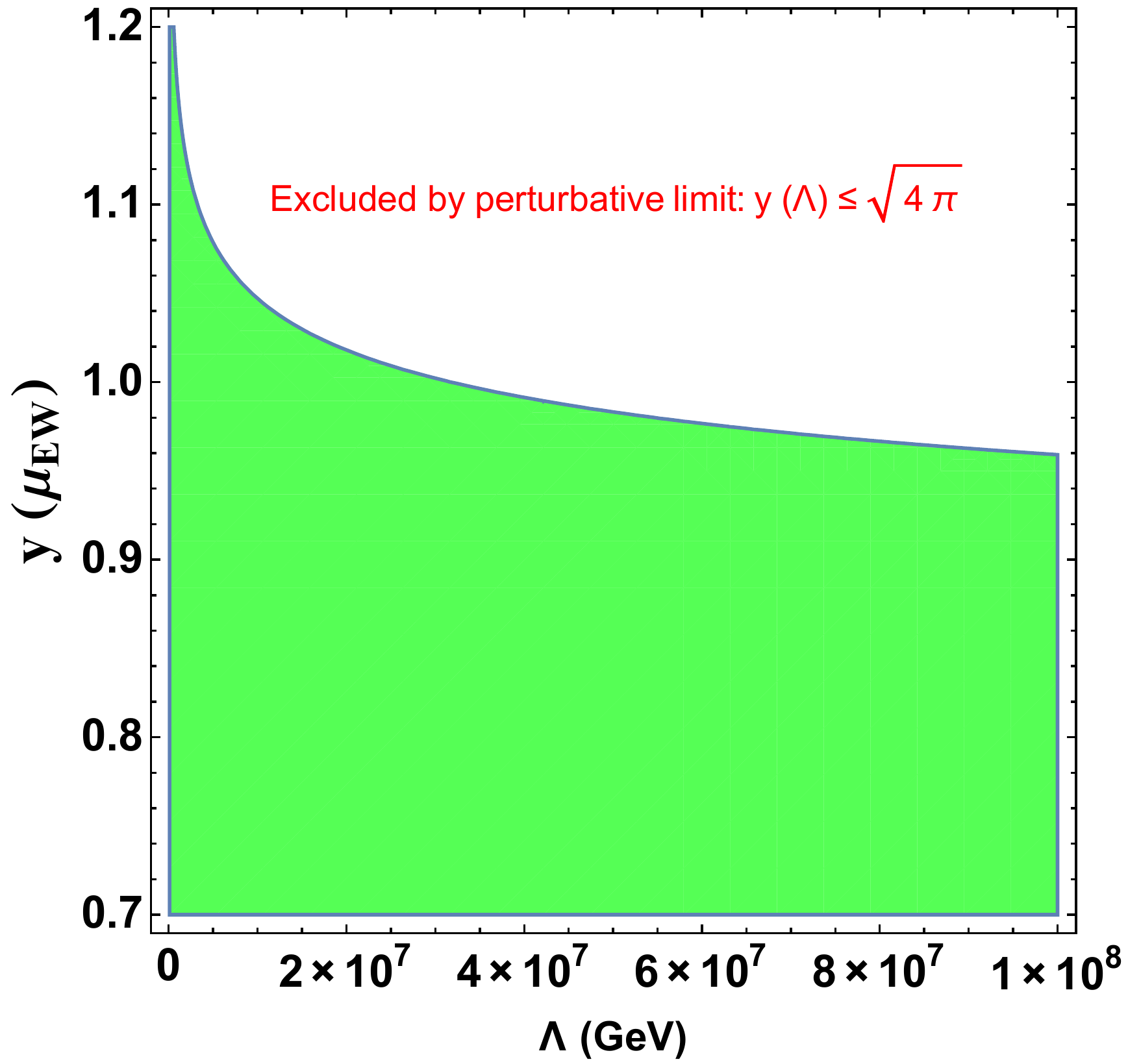}
$$
\caption{Perturbative limit on the $y$ coupling at electroweak scale $y(\mu_{EW})$ assuming $y(\Lambda)=\sqrt{4\pi}$ 
as a function of the cut-off scale $\Lambda$. Green 
region below the curve is allowed by the limit. The range of $\Lambda$ has been chosen from that of our analysis to satisfy neutrino mass limit. }
\label{fig:plimit-y}
\end{figure}

Therefore, in order to keep the $y$ coupling within the perturbative limit near the cut-off scale $\Lambda = M_N$, 
we need to consider suitable value of $y$ at low scale. Below we employ the Renormalisation Group (RG) running of $y$
to evaluate the constraint on it at the electroweak scale.
The RG equation for $y$ is \cite{Bonilla:2015kna} given by:
 \begin{equation}
\frac{dy}{dt}=\frac{6}{16 \pi^2}~y^3,
\end{equation}
where $t=\ln \mu$, with $\mu$ denoting the energy scale. Following our convention used in the analysis, $t_{EW}=\ln(v_{EW})=\ln(174)=5.16$, the solution of above equation will be given by:
 \begin{eqnarray}
\int_{y_0=y(\mu_{EW})}^{y(\Lambda)}\frac{dy}{y^3}=\int_{t_{EW}}^{\Lambda}\frac{6}{16 \pi^2}dt~,\\
\implies y_0=y(\mu_{EW})=1.05, ~~~~ \rm{for}~~~\Lambda=10^7 ~~GeV, \nonumber \\
\implies y_0=y(\mu_{EW})=1.16, ~~~~ \rm{for}~~~\Lambda=10^6 ~~GeV.
\end{eqnarray}
A variation of maximum allowed $y$ at electroweak scale from perturbative limit as a function of cut off scale $\Lambda$ is shown in Fig.~\ref{fig:plimit-y} as well. The range of 
$\Lambda$ has been chosen from that of our analysis to satisfy neutrino mass limit. The green region under the curve provides allowed values $y$.

\section{The case of having $\lambda_1\neq \lambda_2$}
\label{apnd2}
In the previous analysis we have considered $\lambda_1=\lambda_2$ for 
simplicity. However, one can easily extend the analysis by considering 
$\lambda_1 \neq \lambda_2$ for both low and high mass regions. To facilitate
comparison with the previous results, we parametrise $\lambda_1= \kappa 
\lambda_2$ and  briefly 
outline the possible changes. For low mass region, using Eq. (\ref{l1}) - 
(\ref{l2}) and $\lambda_1= \kappa \lambda_2$, we find, 
\begin{align}
 u^2=\kappa v^2\frac{
               m_{H_{\rm light}}^2(1-\cos 2\theta)+m_{h_{\rm SM}}^2 (1+\cos 
               2\theta)
              }
              { 
               m_{H_{\rm light}}^2(1+\cos 2\theta)+m_{h_{\rm SM}}^2(1-\cos 
2\theta)            
              }
\end{align}
and the relevant modified vertex factors are given by   
\begin{eqnarray}
H_{\rm light} h_{\rm SM} h_{\rm SM}&:& 
2\left(\frac{1}{2} v\lambda_{12} c^3_{\theta} - 
3u\lambda_2 c^2_{\theta}s_{\theta} + u\lambda_{12}
c^2_{\theta}s_{\theta}+3v\lambda_1 c_{\theta}s^2_{\theta}- v 
\lambda_{12}c_{\theta}s^2_{\theta}-\frac{1}{2}u\lambda_{12}s^3_{\theta}
\right),\nonumber\\
&=& 
2\left(\frac{1}{2} v\lambda_{12} c^3_{\theta} - 
3u\frac{\lambda_1}{\kappa} c^2_{\theta}s_{\theta} + u\lambda_{12}
c^2_{\theta}s_{\theta}+3v\lambda_1 c_{\theta}s^2_{\theta}- v 
\lambda_{12}c_{\theta}s^2_{\theta}-\frac{1}{2}u\lambda_{12}s^3_{\theta}
\right),\nonumber\\
h_{\rm SM} H_{\rm light} H_{\rm light} 
&:&2\left(\frac{1}{2}u\lambda_{12}c^3_{\theta}+3v\lambda_1
c^2_{\theta}s_{\theta}-v\lambda_{12}
c^2_{\theta}s_{\theta}+3u\lambda_2 c_{\theta}s^2_{\theta}- u 
\lambda_{12}c_{\theta}s^2_{\theta}+\frac{1}{2}v\lambda_{12}s^3_{\theta}
\right),\nonumber\\
&=&2\left(\frac{1}{2}u\lambda_{12}c^3_{\theta}+3v\lambda_1
c^2_{\theta}s_{\theta}-v\lambda_{12}
c^2_{\theta}s_{\theta}+3u\frac{\lambda_1}{\kappa} c_{\theta}s^2_{\theta}- u 
\lambda_{12}c_{\theta}s^2_{\theta}+\frac{1}{2}v\lambda_{12}s^3_{\theta}
\right).\nonumber
\end{eqnarray}
Similarly, using Eq. (\ref{l1high})-(\ref{l2high}) for high mass region, we 
obtain 
\begin{align}
 u^2= \kappa v^2\frac{
               m_{h_{\rm SM}}^2(1-\cos 2\theta)+m_{H_{\rm heavy}}^2 (1+\cos 
               2\theta)
              }
              { 
               m_{h_{\rm SM}}^2(1+\cos 2\theta)+m_{H_{\rm heavy}}^2(1-\cos 
2\theta)            
              },
\end{align}
whereas the modified vertices are given by 
\begin{eqnarray}
h_{\rm SM} H_{\rm heavy} H_{\rm heavy}&:&2\left(\frac{1}{2}v\lambda_{12} 
c^3_{\theta}-3u\lambda_2 c^2_{\theta}s_{\theta}+u\lambda_{12} c^2_{\theta} 
s_{\theta} + 3v\lambda_1 c_{\theta}s^2_{\theta}- v \lambda_{12} c_{\theta} 
s^2_{\theta} - 
\frac{1}{2} u \lambda_{12} s^3_{\theta}\right),\nonumber\\
&=&2\left(\frac{1}{2}v\lambda_{12} 
c^3_{\theta}-3u\frac{\lambda_1}{\kappa} c^2_{\theta}s_{\theta}+u\lambda_{12} 
c^2_{\theta} 
s_{\theta} + 3v\lambda_1 c_{\theta}s^2_{\theta}- v \lambda_{12} c_{\theta} 
s^2_{\theta} - 
\frac{1}{2} u \lambda_{12} s^3_{\theta}\right),\nonumber\\
H_{\rm heavy} h_{\rm SM} h_{\rm SM} &:& 2\left(\frac{1}{2}u\lambda_{12} 
c^3_{\theta}+3v\lambda_1 c^2_{\theta}s_{\theta}-v\lambda_{12} c^2_{\theta} 
s_{\theta}+ 3u\lambda_2 c_{\theta} s^2_{\theta}- u \lambda_{12} c_{\theta} 
s^2_{\theta} 
+\frac{1}{2}v\lambda_{12} s^3_{\theta}\right),\nonumber\\
&=& 2\left(\frac{1}{2}u\lambda_{12} 
c^3_{\theta}+3v\lambda_1 c^2_{\theta}s_{\theta}-v\lambda_{12} c^2_{\theta} 
s_{\theta}+ 3u\frac{\lambda_1}{\kappa} c_{\theta} s^2_{\theta}- u \lambda_{12} 
c_{\theta} 
s^2_{\theta} 
+\frac{1}{2}v\lambda_{12} s^3_{\theta}\right).\nonumber
\end{eqnarray}
These modified expressions for the vev of the additional scalar and Higgs portal 
vertices essentially alters the DM phenomenology.

\begin{figure}[h]
$$
\includegraphics[height=5.4cm]{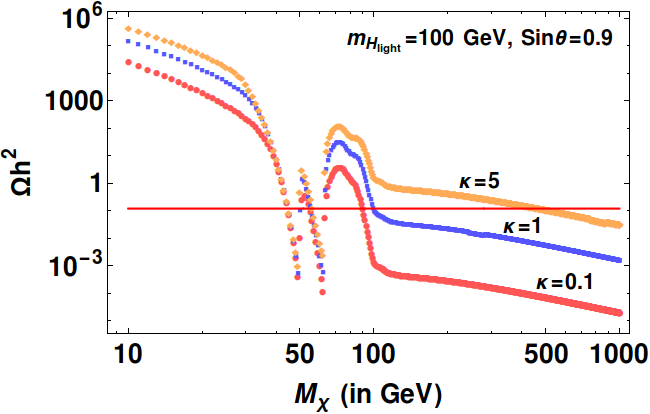}
\includegraphics[height=5.4cm]{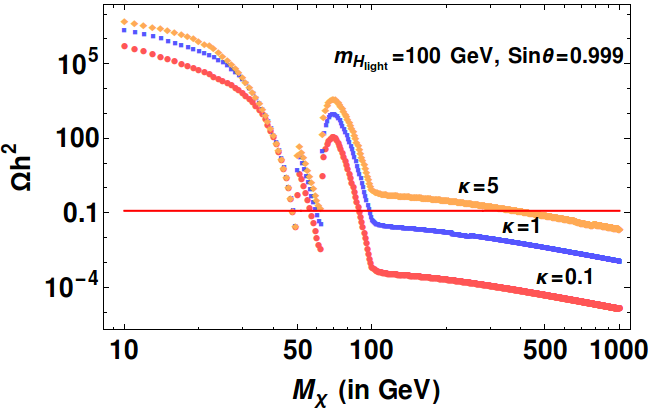}
$$
\caption{Relic density vs DM mass for various $\kappa (=\lambda_1/\lambda_2)$ values 
in the low mass region. Here, orange, blue  and red dotted lines stands for 
$\kappa=5$, $\kappa=1$ and $\kappa=0.1$ respectively.}
\label{fig:clow}
\end{figure}

\begin{figure}[h]
$$
\includegraphics[height=5.4cm]{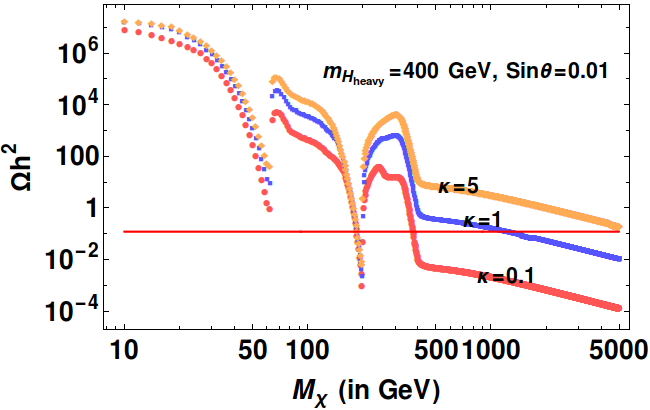}
\includegraphics[height=5.4cm]{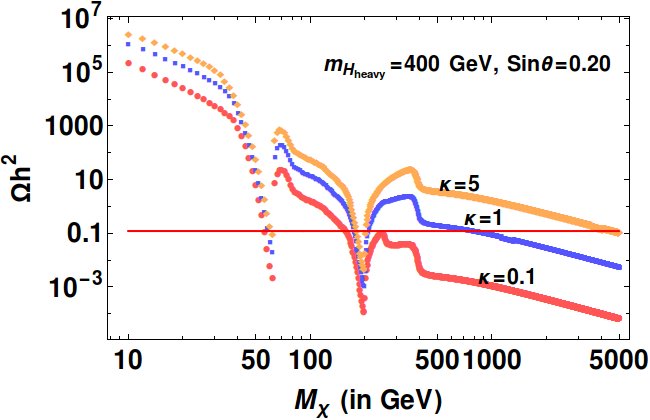}
$$
\caption{Relic density vs DM mass for various $\kappa (=\lambda_1/\lambda_2)$ values 
in the high mass region. Here, orange, blue  and red dotted lines stands for 
$\kappa=5$, $\kappa=1$ and $\kappa=0.1$ respectively.}
\label{fig:chigh}
\end{figure}

Now, in Fig. \ref{fig:clow} and \ref{fig:chigh}, incorporating $\lambda_1 \neq 
\lambda_2$ we have plotted the variation of relic density as a function of the DM 
mass ($M_{\chi}$) for low and high mass regions respectively. We 
observe that as we alter $\kappa$, the Higgs portal coupling also get modified. Here 
a smaller value of effective coupling (when $\kappa=\lambda_1/\lambda_2>1$)  
indicates larger relic density and a larger effective coupling (when 
$\kappa=\lambda_1/\lambda_2<1$) indicates smaller relic density compared to $\kappa=1$ 
case for obvious reasons. This is depicted by the orange ($\kappa=5$), blue ($\kappa=1$) 
and red ($\kappa=0.1$) dotted lines respectively for all panels in Fig. 
\ref{fig:clow} and \ref{fig:chigh}. While in the low mass DM region only 
resonance regions satisfy DM constraints, the effect of this change (in 
terms of $\kappa$) is much more pronounced when the annihilation opens to the other 
light (or heavy) Higgs  for $M_{\chi} > m_{H_{\rm light}}$ (or 
$m_{H_{\rm heavy}}$). This basically lead to a much larger allowed (by both relic 
density and Direct search constraints) parameter space specifically in the high 
DM mass region. 

\section{Higgs invisible decay constraints}
\label{apnd3}

When the DM mass ($M_{\chi}$) is smaller than the SM Higgs mass with $M_{\chi} < m_{h_{\rm SM}}/2 $, then SM Higgs can decay 
to DM and this will contribute to the invisible Higgs decay width. Observations at LHC for the SM Higgs constrains such invisible 
branching fraction as $Br(h\rightarrow inv) < 0.24$ ~\cite{Tanabashi:2018oca}. This can be interpreted in terms of 
invisible decay width as follows:
 \begin{eqnarray}
Br(h \to  inv.) && < 0.24 \nonumber \\
\frac{\Gamma(h \rightarrow inv. )}{\Gamma(h \rightarrow SM) + \Gamma(h \rightarrow inv.  )} && < 0.24,  
\end{eqnarray}
where the Higgs decay width to SM is constrained as $\Gamma(h\to SM)=4.2$ MeV (with mass $m_{h_{\rm SM}}=125.7~ {\rm GeV}$) 
at the LHC. This limits the invisible decay width as, 
\begin{eqnarray}
\Gamma(h \rightarrow inv.  ) < 1.32 ~\rm~ MeV~. 
\end{eqnarray}

\begin{figure}[h]
$$
\includegraphics[height=6.8cm]{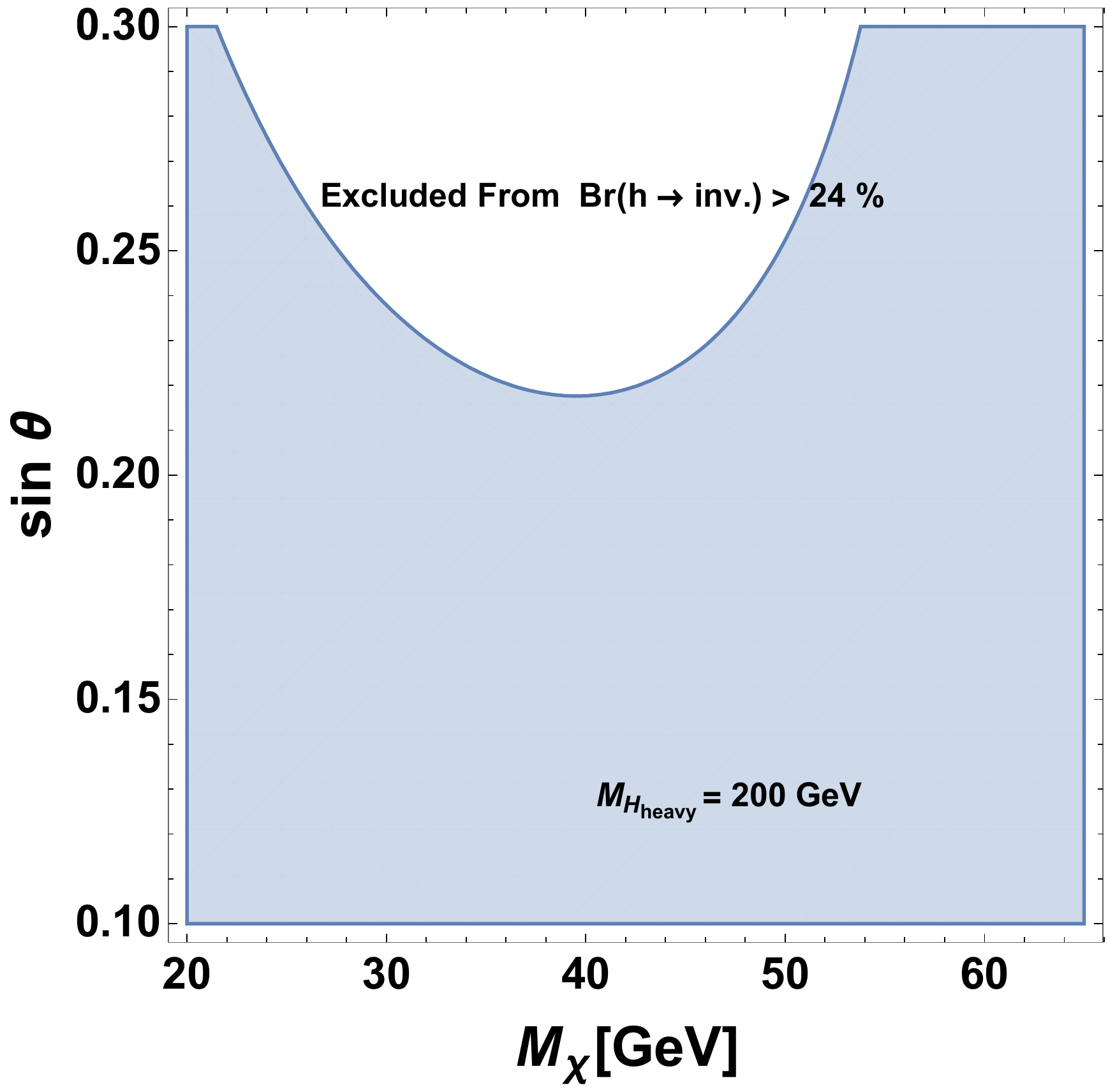}
\includegraphics[height=6.8cm]{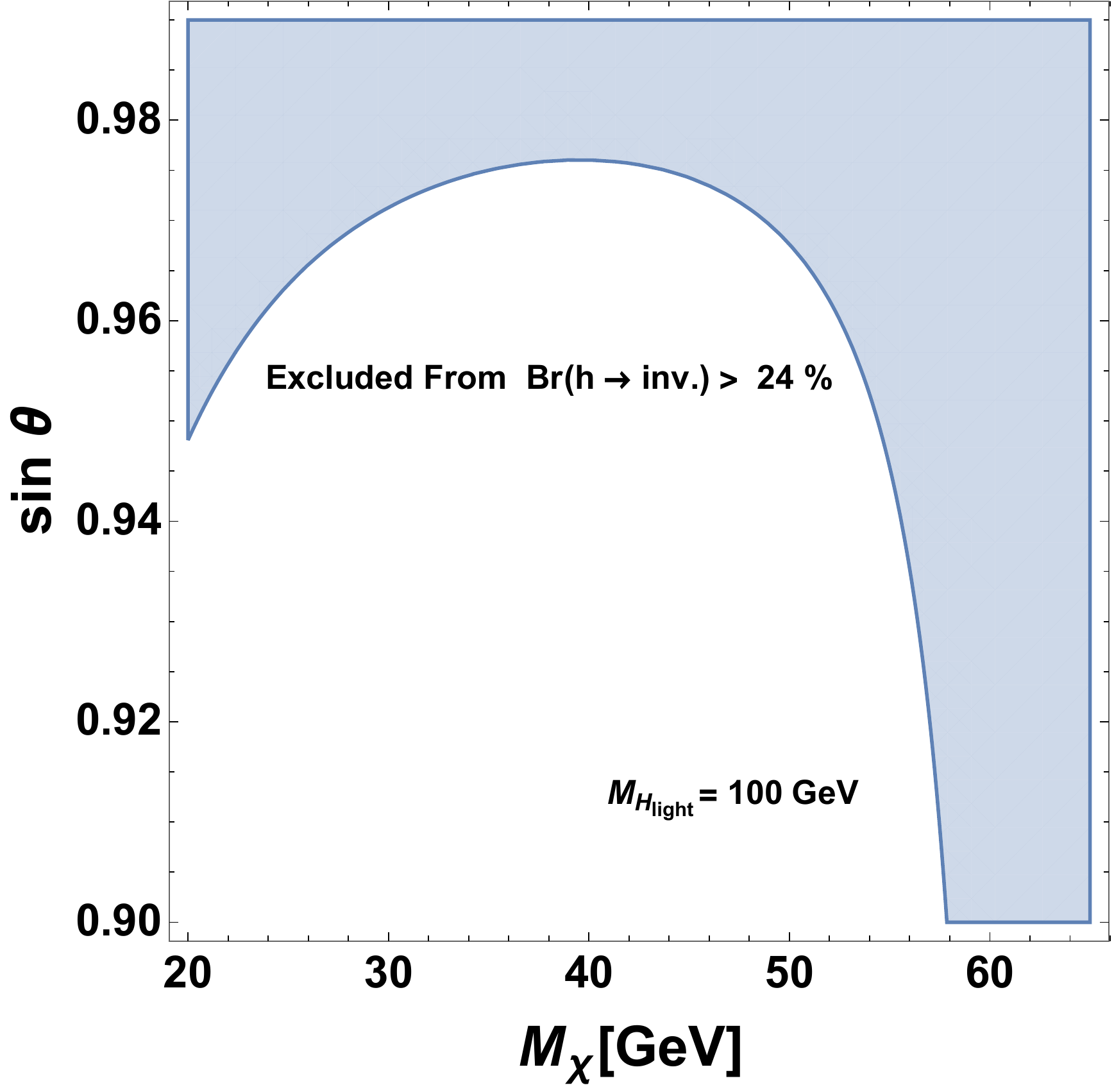}
$$
\caption{Invisible Higgs decay constraint for $M_{\chi} < m_{h_{\rm SM}}/2 $ as a function of scalar mixing angle $\sin\theta$ and DM mass. Left panel shows the case for heavy Higgs 
mass region with $M_{H_{\rm heavy}}=200$ GeV. Right panel shows the case of light Higgs mass with $M_{H_{\rm light}}=100$ GeV. The shaded region is allowed while the white region is discarded.}
\label{fig:hinv}
\end{figure}

The invisible Higgs decay width to DM can be easily calculated:
\begin{eqnarray}
\Gamma_{h \to \chi\chi}= \frac{1}{8 \pi} (y \sin\theta)^2  m_{h_{\rm SM}}~~ \Big(1-\frac{4 M_{\chi}^2}{m_{h_{\rm SM}}^2}\Big)^\frac{3}{2} ~~\Theta(m_{h_{\rm SM}}-  2M_{\chi}) \nonumber \\
\implies \Gamma_{h \to \chi\chi}= \frac{1}{8 \pi} (\frac{M_\chi}{u} \sin\theta)^2 m_{h_{\rm SM}}~~ \Big(1-\frac{4 M_{\chi}^2}{m_{h_{\rm SM}}^2}\Big)^\frac{3}{2} ~~\Theta(m_{h_{SM}}-  2M_{\chi}), 
\end{eqnarray}
where in the above expression, we use the coupling of SM Higgs to DM as $y\sin\theta$ which is valid for the heavy Higgs mass region.
A similar expression for the light mass region can be obtained. Invisible Higgs decay constraint will then put a limit
on the mixing angle $\sin\theta$. In Fig.~\ref{fig:hinv}, we show the limit on $\sin\theta$ in both Heavy Higgs mass region with $M_{H_{\rm heavy}}=200$ GeV in left and light Higgs mass region with $M_{H_{\rm  light}}=100$ GeV. 
The shaded region is allowed while the white region is discarded. The left plot shows that in the high mass region, excepting for the large values of $\sin\theta \sim 0.3$, the small mixing limit is okay with invisible Higgs branching fraction. 
Light Higgs region is more constrained with larger $\sin\theta \sim 0.99$ being allowed.

\end{document}